\documentclass[aps,prb,reprint,amsmath,amssymb,showpacs,floatfix,twocolumn]{revtex4-2}
\usepackage{graphicx}
\usepackage{dcolumn}
\usepackage{braket}
\usepackage{bm}
\usepackage[colorlinks, allcolors=blue]{hyperref}
\usepackage[usenames, dvipsnames]{color}
\usepackage{xcolor}
\usepackage[usenames,dvipsnames]{color}
\usepackage{soul}
\providecommand{\nn}{\nonumber}

\providecommand{\pd}{\partial}

\providecommand{\bv}[1]{\bm{\mathrm{#1}}}

\providecommand{\w}{\omega}
\providecommand{\W}{\Omega}
\providecommand{\q}{\bv{q}}

\providecommand{\p}{\bv{p}}

\renewcommand{\k}{\bv{k}}

\providecommand{\hq}{\hat{q}}
\providecommand{\hk}{\hat{k}}

\providecommand{\vf}{v_F}

\renewcommand{\q}{\bv{q}}

\providecommand{\tpp}{\left(2\pi\right)}

\newcommand{\AK}[1]{#1}
\begin{document}

\title{ Intertwined 
%topology 
geometries 
in collective modes of two dimensional Dirac fermions}

\author{Ankan Biswas}
\affiliation{Physics Department, Ariel University, Ariel 40700, Israel}
%\email{ankanb@ariel.ac.il}
\author{Avraham Klein}
%\email{avrahamk@ariel.ac.il}
\affiliation{Physics Department, Ariel University, Ariel 40700, Israel}

\begin{abstract}

It is well known that the time-dependent response of a correlated system can be inferred from its spectral correlation functions. As a textbook example, the zero sound collective modes of a Fermi liquid appear as poles of its particle-hole susceptibilities. However, the Fermi liquid's interactions endow these response functions with a complex analytic structure, so that this time/frequency relationship is no longer straightforward.
We study how the geometry of this structure is modified by a nontrivial band geometry, via a calculation of the zero sound spectrum of a Dirac cone in two dimensions.
We find that the chiral wavefunctions, \AK{that encode the band geometry,} fundamentally change the analytic structure of the response functions, \AK{which encode its Riemannian geometry}. %giving rise to qualitatively different zero sound spectra from that of a conventional Fermi liquid. 
As a result, isotropic interactions can give rise to a variety of unconventional zero sound modes, that, due to the geometry of the functions in frequency space, can only be identified via time-resolved probes. These modes are absent in a conventional Fermi liquid with similar interactions, so that these modes can be used as a sensitive probe for the existence of Dirac points in a band-structure.

\end{abstract}
\maketitle
%

%\tableofcontents

\section{\label{sec1}Introduction}

Understanding
%The study of 
how a strongly correlated quantum system responds to a strong pulse of electromagnetic radiation combines questions of fundamental physics with exciting prospects 
%raises fundamental questions while offering exciting potential 
for future applications \cite{Giannetti_and_Massimo2016_1}. After being exposed to such a pulse, quantum systems can exhibit strong nonequilibrium dynamics, leading to emergent macroscopic behavior \cite{Gandolfi_2017_2} and metastable states with no equilibrium counterpart, properties that could be crucial for developing quantum switches and devices \cite{RevModPhys.93.041002_3,Nicoletti:16_4}.
Such so-called ``ultrafast'' laser pulses, lasting from attoseconds to picoseconds and spanning frequencies from terahertz (THz) to X-ray\cite{ Sell2008Phase-973_8, Vicario:14_6,RUELLO201521_5,Orenstein2012UltrafastSO_7}, are particularly significant when applied in the low-frequency THz range for quantum materials and devices.
By carefully controlling the wavepacket shape of an incoming pulse, it is possible to probe and manipulate the emergent, low energy collective modes of an interacting quantum system, and measure their time dependent response in a pump-probe configuration.
% In a pump-probe configuration, where a strong pump pulse precedes a delayed weak probe pulse, macroscopic quantum phenomena
Experiments have demonstrated the response of varied quantum phenomena to ultrafast pulses, including magnetism, superconductivity   and many other phenomena \cite{Mitrano2016,Stupakiewicz2021}.

Theoretical descriptions of a pumped system's evolution in the time domain are challenging. One of the reasons of this is that even the simplest out-of-equilibrium observable is a function of \emph{two} times (e.g. center-of-mass and relative time), or of simultaneous time and frequency \cite{Kamenev2011}. In fact, even within a linear response regime, the idea of pump-probe experiments raises a fundamental question on the relationship of time and frequency in interacting systems: \emph{Even in equilibrium}, does the time domain response fully encode the frequency domain response (and vice versa), or do these domains convey different information? In other words, are the frequency and time domain linear responses of an interacting system just naive Fourier transforms of one another, or is there some deeper relationship, potentially with non-local aspects tied to topology?

\begin{figure*}[t]
    \includegraphics[width=0.8\hsize,clip,trim=50 0 435 0]{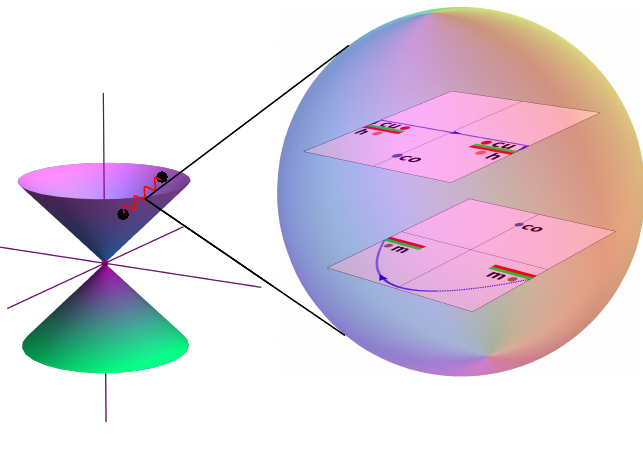}\hspace{0.2\hsize}\llap{\fbox{\includegraphics[width=0.32\hsize,clip,trim=460 0 0 0]{fig1_paper_new1.pdf}}}
    \caption{\label{fig_pictorial}
      Representation of the intertwined geometries of the nontrivial band structure and particle-hole susceptibility $\chi(\q, \W)$. Illustration of the Dirac cone, showing how a particle-hole excitation (red wiggly line) moves an electron by momentum $\q$ and frequency $\W$. Each such process is defined on a Riemann surface (a sphere) of complex $\Omega$. The zero sound collective modes appear as poles on this two-sheeted Riemann surface, of which the top one represents the ``physical'' sheet (whose real axis represents physical frequencies), and the bottom is ``unphysical''. The nontrivial geometry is evident in the gluing of lips of the branch cuts, red to red and green to green. Different colored disks indicate poles representing a variety of collective modes on the two sheets : red disks - conventional underdamped (cu) modes; purple disks - conventional overdamped (co) modes; orange disks - hidden (h) modes, hidden from the physical real axis; and brown disks - mirage (m) modes on the unphysical sheet. The properties of the modes are reviewed in detail in the paper. The time-dependent response is given by a Fourier transform, whose integration contour (blue dashed line) traces over the entire surface, picking up the contributions from the nontrivial geometry. (Inset [in arbitrary units]) The time-dependent response to a sharp pulse, for different angles $\alpha$ between the mode wavevector and its polarization, in a Dirac liquid with a repulsive \emph{isotropic} interaction. The anisotropic frequency and decay rates are a result of the chiral bandstructure, absent in a conventional Fermi liquid. At long times the three responses cross over to a single universal behavior (see Fig. \ref{fig_pl1}). The numerical value used in the inset is $F_0=10$ (see the following sections for details).
    }
\end{figure*}

An efficient way to study this issue is by analyzing the low energy collective excitations of an itinerant fermion system, i.e. a Fermi liquid. The fundamental excitations of a Fermi liquid (in its normal state), are zero sound (ZS) modes.
ZS modes, which are collective excitations from coherent fluctuations of fermions near the Fermi surface, manifest as sharp propagating excitations for repulsive quasiparticle interactions, whereas for attractive interactions, they generally appear as damped resonances originating from quasi-bound states \cite{Abrikosov:107441_24,lifshitz2013statistical_25,Baym1991,Pethick1988,Nozieres1999, Abel1966}. However, as is well-known, the particle-hole response of a Fermi-liquid is non-analytic on the complex frequency plane \cite{PhysRevResearch.1.033134_15}. Elementary theorems of complex analysis imply that the response function is therefore defined a Riemann surface, suggesting that the interactions endow the system with a nontrivial geometry and topological structure \cite{nehari1952conformal_10}. This structure manifests itself through several unconventional zero-sound (ZS) modes \cite{kleinhidden_9}, which we will describe in detail. For example, even in the simplest model of an interacting Fermi liquid, an itinerant paramagnet, weak attraction in the spin channel gives rise to a so-called ``hidden'' mode. This is a sharp propagating mode which is generated by the attraction, rather than repulsion, but which is hidden in the sense that it cannot be detected via standard spectral measurements, due to the nontrivial geometry of the Riemann surface.

The existence of a nontrivial geometric structure arising purely from a Fermi liquid's interactions raises another  question: in itinerant systems with a nontrivial bandstructure geometry and topology, what is the relationship between the two? This relationship is depicted pictorially in Fig. \ref{fig_pictorial}. In this paper, we examine this interplay by studying what is possibly the simplest such case: the ZS spectrum of a Dirac liquid. 
As we shall show, in the low energy limit this relationship is encoded in the particle-hole polarization at finite doping. We computed the zero-temperature one-loop polarization response function,

\begin{flalign}
%\begin{aligned}
\label{eq:Pi-def-intro}
\chi^b_{\mu\nu} (\q,i \W) \approx
% \mbox{Tr} \sum_{\omega_n}
%\sum_{n,m}
%\int \frac{d\w d^2k}{\tpp^3} \bra {u_{\k_-,m}} \tau_{\mu} \ket {u_{\k_+,n}} G_n(\k_+,i \omega_+) \\
%    \bra {u_{\k_+,n}} \tau_{\nu} \ket {u_{\k_-,m}} G_m(\k_-, i \w_-)
-\int \frac{d\w d^2k}{\tpp^3} \bra {u_{\k_-}} \tau_{\mu} \ket {u_{\k_+}}  \bra {u_{\k_+}} \tau_{\nu} \ket {u_{\k_-}}\quad\nn\\
    G(\k_-, i \w_-)G(\k_+,i \omega_+).
  %  \end{aligned}
\end{flalign}

Here, $\epsilon_{\k}$ is the dispersion, $G = (i \w - \epsilon_{\k})^{-1}$ is a fermionic Green's function, $\tau_\mu$ are pseudospin Pauli matrices, and $\ket{u_{\k}}$ is the wavefunction in the conduction band,
% and $\omega, q$ are frequency-momentum vectors. The
and shifted frequencies and momenta are defined as $\omega_\pm = \omega \pm \frac{\W}{2}$ and $\k_\pm = k \pm \frac{\mathbf{q}}{2}$, where $\W$ and $\mathbf{q}$ denote the external frequency and momentum, respectively. 
The existence of the nontrivial form-factors $\bra {u_{\k}} \tau_{\mu} \ket {u_{\k}}$ imprints the band geometry on the polarization. When the bare polarization bubble is renormalized by an RPA sum, this gives rise to collective modes, also ``stamped'' with the band geometry, once the frequency is rotated back to the real axis.

To investigate the interplay of geometries, we identified the collective modes, analyzed their behavior as a function of doping, and compared their characteristics to that of a ``conventional'' Fermi liquid with a parabolic bandstructure and trivial topology. We found, that unconventional ZS modes only appear at finite chemical potential and vanish at the charge neutrality point, implying that the nontrivial geometry of the interaction is a result of Landau damping and the breaking of Lorentz invariance to Galilean invariance. 
%\AK{However, even in the doped case, and in the limit of infinitesimal frequency and momentum transfer, the collective modes dynamics are determined simultaneously from the Riemannian structure due to the interaction and from the chiral band geometry induced by the existence of a Dirac point.}
At finite doping, we found that the chirality of the bandstructure introduces an angular, p-wave like behavior into the off-diagonal ($\mu\neq \nu$) terms in Eq. \eqref{eq:Pi-def-intro}, which creates a host of unconventional ZS modes (see e.g. Fig. \ref{fig_pictorial}). The chiral terms arise from the projection of the wavefunctions onto the pseudospin vector, which is isomorphic to the Berry connection of the system, and indicates that the bandstructure geometry is directly imprinted onto the collective modes. \AK{The modes are qualitatively different from those appearing in a conventional Fermi liquid with a trivial bandstructure, even when accounting for the Riemann surface.} Finally, we did not find any induced change of the genus of the Riemann surface due to the Dirac band structure, indicating that the topological aspects of the system are robust.

\AK{The fact that the ZS modes in a doped Dirac liquid are qualitatively different those of a Fermi liquid, even in the limit of infinitesimal frequency and momentum transfer, when the Dirac point is ``infinitely far away'' from the excitations, means that the two types of geometries are inherently intertwined. In particular, as we will discuss later, the implication is that a knowledge of both structures is necessary for a correct interpretation of the time-dependent response of the ZS collective modes in an interacting system, and conversely, that these modes can be utilized to probe both structures. We demonstrate this further in the Appendix by studying several additional models with parabolic bands, both with and without nontrivial band geometry.}

The rest of this paper is organized as follows. In Sec. \ref{sec2} we introduce our model and review how the Riemannian geometry arises and affects the relationship between time and frequency. In Sec. \ref{sec3} we calculate the ZS collective modes of a Dirac liquid and compare them to that of a conventional Fermi liquid. We make some concluding remarks in Sec. \ref{sec4}.

\section{Model and Theoretical Framework}{\label{sec2} 

\subsection{\label{sec2a}Electronic band structure and interactions }

\begin{figure}[b]
\includegraphics[width=\hsize ]{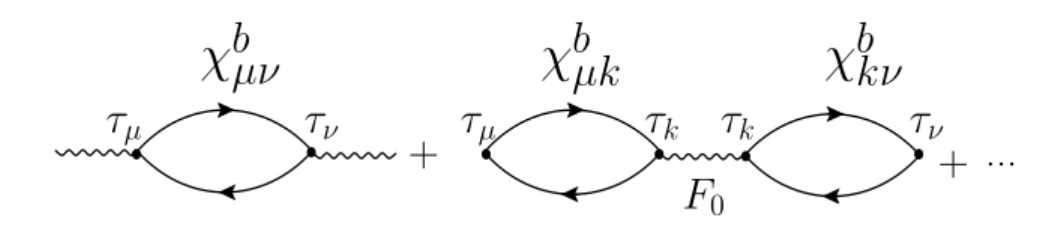}
\caption{\label{fig_crosspol} The RPA sum for the polarization bubbles that determined $\chi_{\mu\nu}$.}
\end{figure}

To study the collective modes in our work, we start with the bare Hamiltonian of a single Dirac cone in the sublattice representation. Neglecting spin degrees of freedom, the Hamiltonian near the Dirac point
%, as illustrated in Fig. \ref{fig_pictorial}(a), 
is expressed as:
%To analyze the polarization bubble and its conversion from sublattice to band form, we start with the Hamiltonian near the Dirac point, expressed as:
\begin{align}
\label{eq:hamil}
    \hat{H} = \sum_{\k} c^{\dagger}_{\k}\left( -\mu \tau_0 + v_F \bv{\tau}\cdot \bv{k}  \right)c_{\k}.
\end{align}
where we use hat notation to denote matrices in the sublattice space. Here $\mu$ is the chemical potential, $\bv{\tau}$ Pauli matrices representing the pseudospin sublattice degree of freedom for the fermions $c_{\k}$, and $v_F$ is the Fermi velocity.
% $\k \to \vf \k$ has been rescaled by the Fermi velocity $\vf$.
This Hamiltonian captures the low-energy fermionic states near the Dirac point, giving rise to the well-known cone structure, as illustrated in Fig. \ref{fig_pictorial}. In physical systems, there are typically several different cones in the Brillouin zone, e.g. six in the case of the tight binding model for graphene \AK{(with an extended Brillouin zone) \cite{altland_simons}}. For most of the calculations in this work, the additional cones just add a numerical prefactor, which we neglect for simplicity. When multiple cones are important, we note it explicitly (see e.g. Sec. \ref{sec3c}).

Inverting the Hamiltonian in Eq.
% Using this Hamiltonian
\eqref{eq:hamil}, the Green's function for the system, on the Matsubara axis, is given by:
\begin{equation}\label{eq:green-bare}
  \hat{G} (\k, i\omega) = -\frac{i (\w - i \mu)\tau_0 - v_F \bv{\tau}\cdot\k}{(\w - i \mu)^2 + v^2_F |\k|^2},
\end{equation}
The bare particle-hole susceptibility is therefore,
\begin{equation}
  \label{eq:chi-bare-sublattice}
  \chi_{\mu\nu}^{b}(\q,i\W) = -\int \frac{d\w d^2k}{\tpp^3} \mbox{Tr}\left[ \tau_{\mu} \hat{G}(\k_+,\w_+)\tau_{\nu} \hat{G}(\k_-,\w_-)\right]
\end{equation}
where the trace is over the sublattice indices. We use Greek subscripts to denote $2+1$ space dimensions, and Latin subscripts for spatial coordinates.
% This Green's function, serving as a foundational element in our analysis of the polarization bubble.
The exact result obtained by diagonalizing the Green's function can be found, along with most computational details, in the Appendix \ref{appA}.

In what follows, we focus on the low energy excitations. We rotate to the real axis via $i\W\rightarrow \W+i\delta$, where $\delta$ is an infinitesimal regularizer for the retarded susceptibility. (Later on we will assume $\delta$ is a small static disorder.)
%and introduce a weak static disorder, which will be crucial to regularize our retarded wavefunctions. 
Finally, we assume the limit
\begin{equation}
  \label{eq:limit}
  v_F|\q|,\W \to 0, s = \frac{\W}{ v_F|\q|} \to \mbox{finite}.
\end{equation}
The implication is that except at the charge neutrality point, which we discuss later, we may always assume the response is dominated by the upper band intraband processes. In that case the full expression in Eq. \eqref{eq:chi-bare-sublattice} simplifies to Eq. \eqref{eq:Pi-def-intro}, where the upper band wavefunctions are
\begin{equation}
  \label{eq:bra-defs}
  \ket{u_{\k}}= \frac{1}{\sqrt{2}}
  \left(\begin{array}{c}
    e^{-i\theta_{\k}} \\  1
  \end{array}\right),
\end{equation}
and $\theta_{\k}$ is the azimuthal angle on the FS. To study the collective modes, we follow the ideas of Fermi liquid (FL) theory. As usual, we assume that the fermionic properties are modified by renormalizing the free propagator in Eq. \eqref{eq:green-bare} with a quasiparticle residue $Z$, $\hat{G}\to Z\hat{G}$ and an effective mass $m^*$, $v_F \to v_F (m/m^*)$, and interact via a static interaction on the FS. Typically, these interactions are decomposed into spin and charge channels.  Here, instead, we consider the interactions decomposed into the sublattice channels \cite{Lundgren2015},
%
% For simplicity, we extend this decomposition to include spin and sublattice channels. We then
and identify the fully renormalized, static, and antisymmetric interactions between low-energy fermions, represented by the vertex
% term
$\Gamma_{ij, kl} (\k-\p)$. The harmonic expansion of $\Gamma_{ij, kl} (\k-\p)$ over orbital moments $l$ is possible for rotationally and SU(2)-invariant systems, such as those we are considering here, and give the sublattice version of Landau parameters $F^{c(s)}_l$, namely

\begin{equation}
	%F_{\alpha \beta, \gamma \delta} = F^c (\k-\p) \delta_{\alpha \gamma} \delta_{\beta \delta} + F^s (\k-\p) \bv{\tau}_{\alpha \gamma} \cdot\bv{\tau}_{\beta \delta},
    \Gamma_{ij, kl} = \nu_F^{-1} F^c (\k-\p) \tau_0^{ik}\tau_0^{jl} + \nu_F^{-1} F^s (\k-\p) \bv{\tau}^{ik} \cdot\bv{\tau}^{jl},
\end{equation}
%
%
% Here,
where
\begin{equation}
\label{eq:f0expnd}
F^{c(s)} (\k-\p) = F^{c(s)} _0 + 2 \sum_{l=1}^{\infty} F^{c(s)} _l \cos l (\theta_{\k}-\theta_{\p})    
\end{equation}
and
%Here, $F^{c(s)}_l = \nu_F F^{c(s)} _l$, where 
$\nu_F =  \frac{\mu}{v_F^2} Z^2 \left(\frac{m^*}{m}\right)$, 
%with $D_F$ the density of states at the Fermi energy,% $Z$ is the quasiparticle residue and $m^*$ is the quasiparticle effective mass, 
where
%and 
$\theta_{\k}, \theta_{\p}$ are the azimuthal angle of $\k$, $\p$.

As a starting point for the analysis, let's compute the (retarded) density-density reponse $\chi^R_{00}$. 
We sum over the sublattice degrees of freedom, go through the algebra and find \cite{kleinhidden_9}, for finite $\mu$,

\begin{equation}
  \label{eq:Pi-final}
  \chi^b_{00} (s)= \frac{\mu}{v^2_F} l_0(s),% \left(1 - \frac{q^2}{8\mu^2}\right)
\end{equation}
where
\begin{equation}
 % \begin{aligned}
  l_0 (s)=% 1+ \frac{i s}{\sqrt{1-(s+i\delta)^2} - \delta}. 
   1+ \frac{i s}{\sqrt{1-s^2}}. 
   % \qquad \qquad  s=\frac{\w}{\vf q}.
  \label{eq:lo}
  % \end{aligned}
\end{equation}

Up to the global prefactor, this is precisely the same response as for a conventional FL \cite{kleinhidden_9}. 

At the charge neutrality point (CNP), we have \cite{Wunsch_2006_11, hwang2007dielectric_31},
\begin{equation}
  \label{eq:cnp}
  \chi^{CNP}_{00} (\q,s)= \frac{
  |\q|}{32 v_F} 
  %\textcolor{Green}{ \frac{-1}{\sqrt{s^2-1}}\, .}
  \frac{1}{\sqrt{1-s^2}}.
\end{equation}

Viewed as a function of complex $s= \frac{\W}{\vf |\q|}$, $\chi_{00}(s)$ has branch points at
\begin{equation}
  \label{eq:branches}
  \pm s_{b} =\pm 1 - i \delta,
\end{equation}
where, as we recall, $\delta$ is an infinitesimal regularizer that insures $\hat{\chi}$ is analytic in the upper half plane. 

It is these branch points
% and branch cuts
that introduce the nontrivial geometry and topology into the structure of $\hat{\chi}$, as we now discuss.

\subsection{\label{sec2b}The analytic structure of the collective modes}

Let us now turn to discuss the ZS collective mode spectrum and its associated algebraic structure. The ZS modes are the poles of the quasiparticle susceptibility, computed from the RPA sum of the polarization bubbles and the interaction vertex, in the limit of vanishing momentum and frequency, see Eq. \eqref{eq:limit}. For a generic interaction the computation of the sum can be quite involved. Here, to clarify the algebraic structure we will consider the simplest limit of a single Landau parameter $F_0^{(s)}= F_0$. We choose to use the pseudospin channel rather than the charge channel to avoid issues of the Coloumb interaction and plasmons \cite{lucas2018electronic_12}. As we shall see, for the Dirac case the existence of  chiral wavefunctions qualitatively changes the collective mode behavior from, e.g. a spin fluctuation channel.

For the single Landau parameter, the sum is straightforward and given by Fig. \ref{fig_crosspol}. The vertex $F=F_0 \tau_m\tau_m, m = x,y$, and the sum is,
\begin{equation}
  \label{eq:RPAsum}
  \chi^{qp}_{\mu\nu} = \chi^b_{\mu\nu} - F_0 \chi^b_{\mu k}\chi^b_{k\nu}+ \cdots = \left[(\hat{1} + F_0 \nu_F^{-1}\hat{\chi}^b)^{-1}\hat{\chi}^b\right]_{\mu\nu},
\end{equation}
where the superscript ``qp'' denotes the ``quasiparticle'' susceptibility \cite{PhysRevB.64.195109_13, PhysRevB.97.165101_14, PhysRevResearch.1.033134_15, beal1994zero_17, anderson2011polarization_18, li2012fermi_19, zyuzin2018dynamical_20, khoo2019shear_21, lucas2018electronic_22, torre2019acoustic_23}. Henceforth we supress the ``qp'' notation.
To properly understand the pole structure, we must recall that the bare susceptibility \cite{Abrikosov:107441_24, lifshitz2013statistical_25, PhysRev.140.A1869_26}, and hence quasiparticle susceptibility as well, includes a square-root singularity and has two branch points in the lower half-plane, right below $s=\pm 1$, see Eqs. (\ref{eq:Pi-final})-(\ref{eq:branches}).

As a result, $\hat{\chi}$ in
% \eqref{eq:Pi-final}
Eq. (\ref{eq:RPAsum}) is not defined on a complex plane but rather on a two sheeted Riemann surface. One sheet which we denote the ``physical'' sheet includes the physical real $s$ axis and is analytic in the upper half plane. The other sheet, the ``unphysical'' one, does not need to be analytic in the upper half plane. These sheets are depicted in the illustration in Fig. \ref{fig_pictorial}, showing how poles of $\hat{\chi}$ on the surface can exist in three of the four available half-planes. To construct the surface, we ``cut'' the sheets along branch cuts, chosen here to run from $\pm s_b ... \pm \infty -i \delta$, where we recall that on the sphere all points at infinity converge. The genus of the surface, which in our simple case is just zero (a closed sphere), and the number of poles on it, are topological invariants.

As discussed in the Introduction, the nontrivial Riemann structure has a profound impact on the time-dependent response. For an instantaeous pulse exciting the FL, the response will be given by the Fourier transfer of $\hat \chi$,
\begin{equation}
 \label{eq:time}
\hat{\chi} (\q,t) = \int_{-\infty}^{\infty} \frac{ds}{2 \pi} e^{-i s v_F |\q| t}\hat{\chi}(s).
\end{equation}

As seen in Fig. \ref{fig_pictorial}, the complex integration contour runs over the surface, and does not have to be chosen strictly on the physical sheet. This implies that the time-dependent response can be dominated by poles that are not ``conventional'', in the sense of being situated below the physical $s$ axis and which show up as the well known sharp Lorentzians or broad damped peaks in the spectral function.

For a concrete example of this, let's imagine the interaction of Eq. (\ref{eq:f0expnd}) is in fact that of a conventional parabolic FL, such that $F_0$ from Eq. (\ref{eq:RPAsum}) is just an isotropic spin interaction (for an itinerant ferromagnet), and we are interested in the spin response $\chi_{ij}$. The sum is completely straightforward such that,
\begin{equation}
    \chi_{ij} \propto \delta_{ij}
\end{equation}and the poles are just the well-known poles of a FL \cite{Abrikosov:107441_24,lifshitz2013statistical_25} ZS mode ,
\begin{equation}
  \label{eq:spin-l0-mode}
  1 + F_0 l_0(s) = 0.
\end{equation}
Since $l_0(s=0)=1$, Eq. (\ref{eq:spin-l0-mode}) is a manifestation of the Pomeranchuk criterion, requiring $-1 < F_0$. For a repulsive interaction ($F_0 > 0$) one obtains the well-known ZS propagating modes, \cite{lifshitz2013statistical_25}, \cite{PhysRevResearch.1.033134_15}, \cite{PhysRevB.97.165101_14}}. For a weak attractive interaction $-1/2 < F_0 < 0$, one obtains \emph{hidden modes} that reside just below the branch cut on the physical sheet \cite{PhysRevResearch.1.033134_15,kleinhidden_9}, see the orange disks in Fig. \ref{fig_pictorial}. These are poles at positions
\begin{equation}
  \label{eq:hidden-l0}
  s_h = \pm (1+\sigma_h) - i Q \delta, 
\end{equation}
with $Q > 1, \sigma_h > 0$. The poles cannot be seen in the spectral function due to the branch cut, but completely determine the time-dependent response function, which can be shown to be
\begin{equation}
  \label{eq:chit-hidden-l0}
  \hat{\chi}(t) \propto
  \left\{\begin{array}{ll}
    \frac{\cos(v_F|\q| t + \pi/4)}{(v_F|\q| t)^{1/2}}, & \sigma_h v_F|\q| t \ll 1\\
     \frac{\cos(v_F|\q| t - \pi/4)}{\sigma_h (v_F|\q| t)^{3/2}},  & \sigma_h v_F|\q| t \gg 1
  \end{array} .\right.
\end{equation}

It is clear, that we can expect the existence of a chiral bandstructure to have some influence on the geometric structure of $\hat{\chi}$. In the next section, we will find the analytic structure of $\hat{\chi}$ for the Dirac liquid, identify the unconventional modes that arise for the liquid, and compare it to the case of a conventional FL.

\begin{figure}[t]
\includegraphics[width=\hsize ]{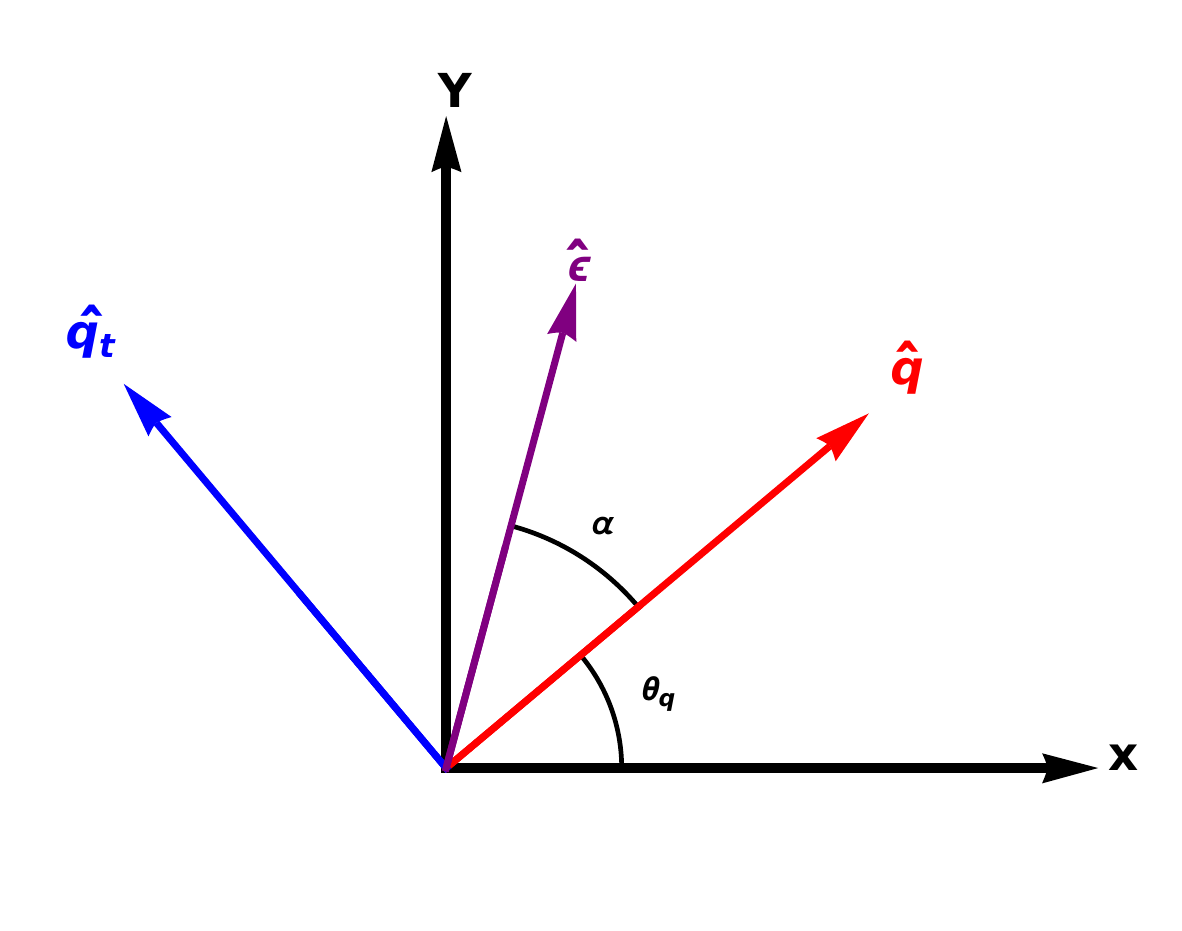}
\caption{\label{fig_vectors} The basis for describing the longitudinal response $\chi(\alpha)$. The components transverse to $\hq$ is $\hat{q_t}$.}
\end{figure}

\section{\label{sec3}Collective modes in a Dirac Liquid}

In the previous section, we presented the analytical structure of the collective modes, which lead to the time-dependent response function $\hat{\chi} (t)$, as given in Eq. \eqref{eq:time}. As discussed, the key ingredients determining this function are the positions of the poles on the Riemann surface, which determine the nature of the unconventional collective modes. In this section, we analyze the response functions and the positions of their poles on the Riemann surface. Then we consider their impact on spectral and time-dependent signatures. Although some of the intermediate results in our analysis have appeared before \cite{kleinhidden_9}, we include them here to clarify the discussion.

We begin by considering the analytic structure of the bare bubble $\hat{\chi}^b$, Eq. \eqref{eq:chi-bare-sublattice}. As noted in Sec. \ref{sec2a}, at the charge neutrality point, $\chi^b \sim |\q|$, implying that zero-sound modes are strongly suppressed. Furthermore, even for finite $\q$, it can be verified that the modes are conventional ZS modes \cite{PhysRevB.75.205418_27}. The reason for this is just that at the charge neutrality point the spectrum consists entirely of interband transitions, and these are not sensitive to the small $\W,\q$ singularity of the polarization. We therefore henceforth limit ourselves to the doped case $\mu>0$. Because we are considering a low energy limit, see Eq. \eqref{eq:limit}, we are always in the regime where to zeroth order in $\w,|\q|$ (but not $s$!), only the upper band intraband contribution is relevant. This justifies our limiting the analysis to the upper band, see Eq. (\ref{eq:bra-defs}). More complete expressions, as well as detailed calculations of the various bubble diagrams, can be found in Appendix \ref{appA}.

For reasons that will become clear shortly, it is more convenient to start by analyzing the pseudospin response $\chi_{\mu\nu}=\chi_{ij}$ with $i,j = x,y$, rather than the density-density $\mu=\nu=0$ channel (it can be checked that in a Dirac system $\chi_{z\mu} = 0$). 
As is evident from Eq. \eqref{eq:Pi-def-intro}, these have a nontrivial dependence on the form factors induced by the chiral wavefunctions, namely
\begin{equation}
\label{eq:Pi1-formfactor}
    \bra{u_{\k}}\tau_0\ket{u_{\k}} = 1,~~ 
    \bra{u_{\k}}\bv{\tau}\ket{u_{\k}} = \left(\begin{array}{c}
          \cos\theta_{\k}\\\sin\theta_{\k}\\0
    \end{array}\right)
\end{equation}

It is,
\begin{align}
 \label{eq:Pi1-final_matrix}
  \chi^b_{ij} (\hq, s) &= \frac{\mu}{8\pi v_F^2} \left[f_2(s) \delta_{ij}  + \left(f_1(s)-f_2(s)\right) \hat{q}_{i} \hat{q}_{j}\right ]  \\
  &= \frac{\mu}{8\pi v_F^2} \left[ f_1(s) \hq \hq + f_2(s) \hq_t\hq_t\right].\label{eq:anglechi}
\end{align}
Here, 
% In this formulation,
$\delta_{ij}$ represents the identity matrix, while $\hq = (\cos\theta_q, \sin\theta_q)$ denotes a longitudinal unit vector for the momentum, and $\hq_t = \hat{z}\times\hq$ is the transverse one. $f_1(s)$ and $f_2(s)$ are Lindhard functions, analogous to the p-wave, angular momentum $\ell=1$ channel, Lindhard functions in a FL \cite{wu2007fermi_28,kleinhidden_9}:
%\AK{CITE WU SUN FRADKIN ZHANG 2007}
\begin{align}
 \label{eq:f1f2_1}
f_1 (s)&= 1 + 2 s^2 \left(\frac{\sqrt{1-(s+i \delta)^2}+ i(s+i\delta)}{\sqrt{1-(s+i \delta)^2} - \delta}\right)\, ,\\
f_2(s) &= 1- 2s \left(s+i\delta - i\sqrt{1-(s+i\delta)^2}\right)\, .\label{eq:f1f2_1_2}
\end{align}
The appearance of the $\ell=1$ functions is a direct consequence of the chiral form-factors in Eq. \eqref{eq:Pi-def-intro} \cite{Raghu2010}, and as we shall see, they give rise to the qualitative differences between the response of a Dirac and Fermi liquid. Such a difference is missing in the bare form of $\chi_{00}^b$, which as noted above is identical to that of a FL, see Eq. \eqref{eq:Pi-final}. \AK{Here, and henceforth, we promote the infinitesimal $\delta$ to a weak static disorder, $\delta = \gamma/(v_F|\q|)$, where $\gamma$ is some small static scattering rate. This is done to more accurately model the motion of the poles in the complex planes, by summing up the disorder ladder diagrams. It can be shown that for a conventional FL, the diffusion pole becomes relevant only in the $\delta \gg 1$ limit and predominantly in the $\ell = 0$ channel \cite{PhysRevResearch.1.033134_15,kleinhidden_9}. Hence, we may expect disorder effects, e.g. weak localization, to be negligible. We also implicitly assume that temperature is low enough to neglect thermal scattering, but high enough that pairing does not set in (assuming such interactions exist in our system).}

We now turn to the fully renormalized $\chi_{ij}$, see Eq. \eqref{eq:RPAsum}. We consider an excitation with a polarization (see Fig. \ref{fig_vectors}),
\begin{equation}
\hat{\epsilon}= \cos\left(\alpha\right) \hq + \sin(\alpha) \hat{q_t},\label{eq:polarize}  
\end{equation}
which can be the result of e.g. a pulse exciting a phonon in the system \cite{castro2009electronic_29}. Since we are interested mainly in the nontrivial geometric properties in this paper, we do not explicitly include the phonon dynamics in our analysis, considering for simplicity a pure electronic mode. To be concise, in what follows we describe the case of a linearly polarized experiment for both incoming and outgoing beams. In that case the RPA sum is straightforward, and yields

\begin{equation}
\begin{aligned}
\label{eq:anglechil}
  \chi(\alpha) &= \hat{\epsilon}\cdot\hat{\chi}\cdot\hat{\epsilon} \\
  &= \cos^2(\alpha) \left(\frac{ f_1(s)}{1 + F_0 f_1(s)}\right) + \sin^2(\alpha) \left(\frac{f_2(s)}{1 + F_0 f_2(s)}\right)\, ,
\end{aligned}
\end{equation}
Eq. \eqref{eq:anglechil} describes two modes, one longitudinal and one transvese, that can be accessed by varying the angle of $\hat{\epsilon}$ compared to $\hq$ (note that any underlying lattice symmetries are irrelevant). This response is deceptively similar to what one might find for a FL interacting via a vector-like mode with a p-wave form factor \cite{kleinhidden_9}, but the physics is completely different. We stress that the interaction (quantified by $F_0$) here is completely isotropic, see Eq. \eqref{eq:f0expnd}, and the angular dependence is purely a result of the chiral band structure. The response also maintains rotational invariance, differently from what would occur due to e.g. a lattice-induced anisotropy. Hence, the collective modes we describe below are what should be expected of any interacting Dirac system, at small enough energies.

We now proceed to find the poles of $\chi(\alpha)$, which are given by the characteristic equation
\begin{equation}
	\label{eq:zero-pole}
\left(1+ F_0 f_1(s)\right) \left(1+ F_0 f_2(s)\right) = 0\, .
\end{equation}
The evolution of these poles are precisely the evolution of the Riemann surface.

\subsection{\label{sec3a}Solution for the poles}

\begin{figure*}[t]
\includegraphics[width=\hsize]{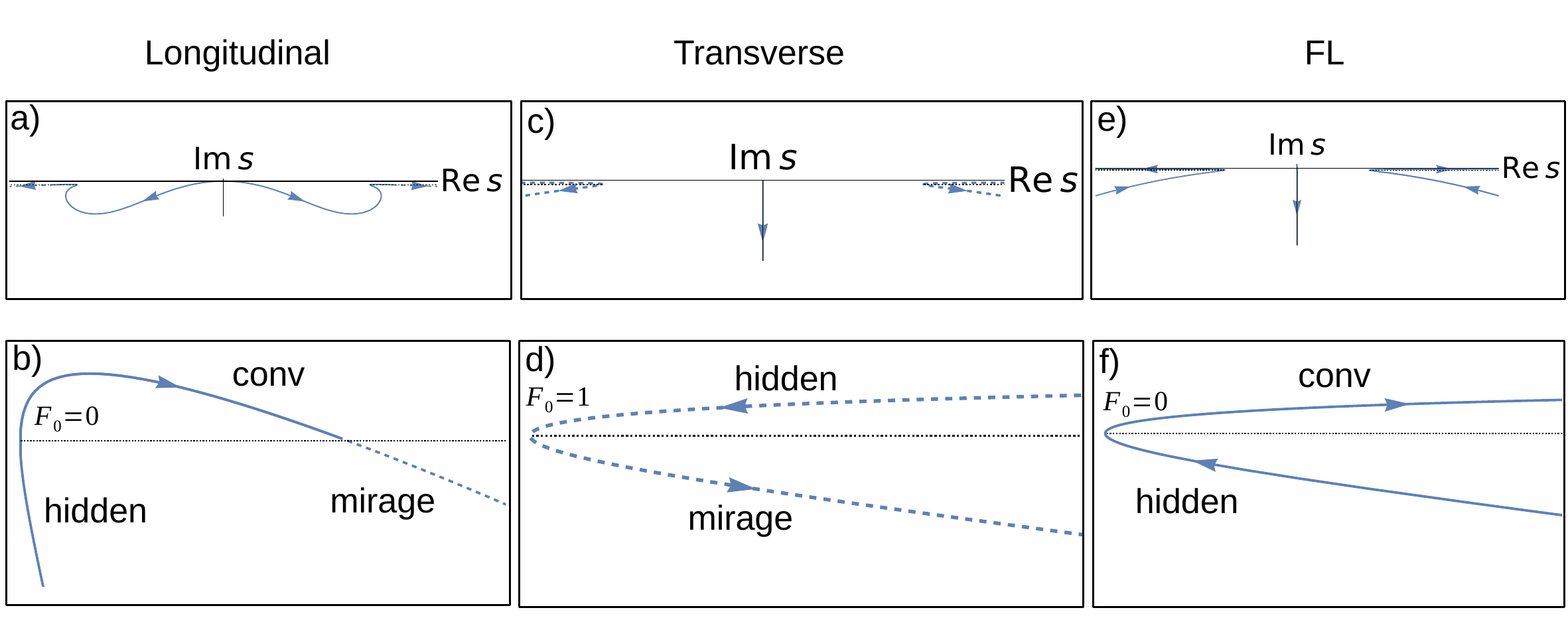}
\caption{\label{fig_pl}Evolution of the poles of $\chi (s)$ (solutions of Eq. \eqref{eq:zero-pole}) in the complex plane as a function of the interaction parameter $F_0$.
%for finite disorder ($\delta =.025$). 
Solid blue lines represent the movement of poles on the physical Riemann sheet, while dashed blue lines indicate motion on the unphysical Riemann sheet. The black horizontal dotted lines denote the branch cuts of $\chi (s)$ at $\mbox{Im s} = -\delta$ and $|\mbox{Re s}| > 1$. Arrows indicate the direction of pole motion as $F_0$ increases from $-1$ to $\infty$. The upper panels depict an overview of the pole trajectories, while the lower panels provide a zoomed-in view of the branch point region at $s=1-i\delta$, allowing one to clearly distinguish why there are conventional,
%(orange circles), 
hidden,
%(dashed circles), 
and mirage modes
%(green circles) 
in Fig. \ref{fig_pictorial}. Note how for the Dirac longitudinal (b), and FL single mode (f), the pole moves on the physical sheet from below to above the branch point, whereas for the Dirac trasverse mode (d), it moves from above to below on the unphysical sheet. For large $F_0 > 0$, both longitudinal and transverse modes are mirage ones: they reside below the cut on the unphysical sheet. A detailed explanation of these modes and the critical values of $F_0$ can be found in Sec. \ref{sec3a}. The disorder strength in these figures, as well as all the following figures, is $\delta = 0.025$.
}
\end{figure*}

The characteristic Eq. \eqref{eq:zero-pole} admits two solutions and six poles in Riemann surface: either $(1 + F_0 f_1 (s)) = 0 $ or $(1 + F_0 f_2(s)) = 0$, unless both terms vanish simultaneously. The nature and trajectory of the resulting poles depend sensitively on the interaction parameter $F_0$ and disorder strength $\delta$, revealing two distinct unconventional modes: hidden and mirage poles.

\subsubsection{Poles dictated by the longitudinal mode \texorpdfstring{$f_1$}{f1}}

The first set of poles arises from solving
\begin{equation}
1 + F_0 f_1(s) = 0,
\end{equation}

which is associated with 4 poles.
For attractive interactions in the range $-1<F_0<0$, the small disorder is irrelevant, and it is enough to analyze Eq. \eqref{eq:f1f2_1} in the clean limit, where the analytic form of $f_1$ is the same on both sheets, namely
\begin{equation}
  \label{eq:f1-clean}
  f_1 (s)= 1 + 2 s^2\left(1+\frac{i s}{\sqrt{1-s^2}}\right).
\end{equation}
Near the Pomeranchuk transition $F_0 \simeq -1$, two poles reside on the \emph{physical} sheet, right \emph{below} the origin, and two on the \emph{unphysical} sheet, right \emph{above} the origin. They move in the complex plane as $|F_0|$ decreases, see Fig. \ref{fig_pl}a (the motion of the unphysical poles is a mirror image of the physical ones). In order to properly analyze the pole behavior when its real part is greater than 1, see Eq. (\ref{eq:branches}), Eq. (\ref{eq:f1-clean}) must be properly extended, and has the form,
\begin{align}
  \label{eq:f1-long-ext}
  f_1 (s)&= 1 + 2 s^2 \left(\frac{\mp i \sqrt{(s+i \delta)^2 - 1}+ i(s+i\delta)}{\mp i\sqrt{(s+i \delta)^2- 1} - \delta}\right) \nn\\
  &= 1 + 2 s^2 \left(\frac{ \sqrt{(s+i \delta)^2 - 1}\mp(s+i\delta)}{\sqrt{(s+i \delta)^2- 1} \mp i \delta}\right),
\end{align}
where the sign is \emph{negative above} the branch cut and \emph{positive below} it on the \emph{physical} sheet, and the converse on the unphysical sheet. 

At a critical value (for $\delta \to 0)$
\begin{equation}
  \label{eq:f0-h}
  F_0^h = -\frac{1}{9}
\end{equation}
the poles touch the branch cut on the physical (unphysical) sheet from below (above), and move towards $\pm s_b (\pm \infty)$. Weak finite disorder induces a slight curvature to the motion of the poles on the complex plane. For positive $F_0$ the poles on the physical sheet move above the branch cut, and proceed to evolve till a second critical value
\begin{equation}
F_0^{m} = \frac{3}{5}.\label{eq:f0-m}
\end{equation}
At this point the poles move through the branch cut to the unphysical sheet, see Fig. \ref{fig_pl}b. The other two poles remain on the unphysical sheet throughout.

\subsubsection{Poles dictated by the transverse mode $f_2$}

% To explore the emergence of hidden modes, we first analyze the solution of
The second set of poles is given by the solution of the transverse mode equation
\begin{equation}
  \label{eq:f2-pole}
  1 + F_0 f_2(s) = 0,
\end{equation}
which is associated with 2 poles. 
% in the clean limit ($\delta \to 0$), where
Similarly as for the longitudinal mode, for attractive interactions in the range $-1<F_0<0$ the analytic form of $f_2$ is the same on both sheets, obeying (for $\delta\to 0$)

\begin{equation}
f_2(s) = 1 - 2 s^2 + 2i s \sqrt{1-s^2}. 
\end{equation}
For $F_0 \simeq -1$ the poles reside just below (above) the real axis on the physical (unphysical) sheet, and as $|F_0|$ decreases, the poles are pushed deeper into the complex plane, moving on the imaginary axis, see Fig. \ref{fig_pl}c. For $|F_0|\to0$ they asymptotically approach $s = \mp i \infty$, which we remind the reader, are connected, since $|s| =\infty$ is a single point connecting both physical and unphysical sheets.
%on the physical Riemann sheet, as illustrated in Fig. \ref{fig_pl}(b). Notably, for $F_0 = 0+$, the pole disappears from the physical sheet.
For $F_0 > 0$ the poles bifurcate to the \emph{unphysical} sheet, above the branch cut but below the real axis, and as $F_0$ increases they move inwards towards the branch point, with the appropriate form of $f_2$ being,
\begin{equation} 
\label{eq:f2-branch}
f_2(s) = 1 - 2s \left(s+i\delta \mp \sqrt{(s+i\delta)^2 -1}\right),
\end{equation}
again with  the sign \emph{negative above} the branch cut and \emph{positive below} it on the \emph{physical} sheet, and the converse on the unphysical sheet.

As the interaction strength increases, the pole evolves continuously across the complex plane, transitioning from above the branch cut to below it at the critical value
%(for $\delta\to 0$),
(for any $\delta \ll 1$),
\begin{equation}
  \label{eq:F0-mh}
  F_{0}^{hm} = 1 .
\end{equation}
At this point (see Appendix \ref{appB}), the pole abruptly becomes visible on the physical real axis, see Fig. \ref{fig_pl}d, but it remains on the unphysical sheet. %The detailed analytical calculation can be found in Appendix \ref{appB} %As long as $F_0 < F_{mh}$, the pole represents a hidden mode, since it cannot

In Eqs. (\ref{eq:f0-h}), (\ref{eq:f0-m}) and (\ref{eq:F0-mh}), we chose the notations ``h'', ``m''. These denote respectively hidden and mirage modes, which are the two types of unconventional ZS modes that appear in these systems, and whose behavior we describe below.

\begin{figure*}[t]
\includegraphics[width=\hsize]{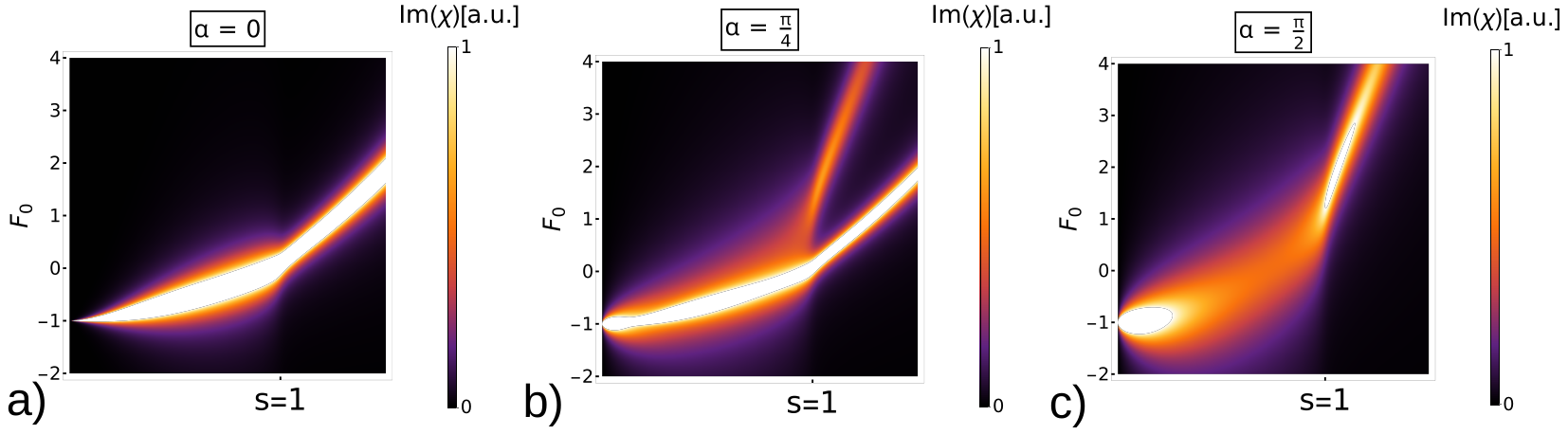}
\caption{\label{fig_dens} The spectral function $\mbox{Im} \chi (\alpha, s)$ (in arbitrary units)
%, with a dimensionless coupling constant $\delta =.025$ characterizing finite disorder and polarization angle $\alpha$, is shown 
for different values of $\alpha$. a) The evolution of the longitudinal mode ($\alpha =0$), b) a mixed case ($\alpha=\frac{\pi}{4}$), and c) the transverse mode ($\alpha = \frac{\pi}{2}$). The Pomeranchuk instability is clear in all figures for $F_0 = -1$. As $F_0$ evolves to more positive values, in all cases, the mode goes from being damped (intrinsically broad as a function of $s$) and below the particle-hole threshold $(s<1)$, to underdamped (limited only by disorder) at $s > 1$. However, the underdamped width still grows with the interaction strength, in stark contrast to the usual FL ZS behavior, as these are all mirage modes. In addition, panel c) shows the transverse mode disappearing and then reappearing above the particle-hole threshold, creating an illusion of two distinct modes. In actuality, the mode just became hidden in the middle. See Fig. \ref{fig_pl3} for the same behavior plotted as a function of angle. }
\end{figure*}

\subsection{\label{sec3b}Spectral Signatures and Real-Time Evolution of Hidden and Mirage Modes}

Having established the pole structure of $\chi(\alpha, s)$, we now explore how these unconventional modes manifest in the spectral function. 

The main properties of the spectral function are depicted in Fig. \ref{fig_dens}. For attractive interactions, there are two damped modes, both of which become unstable at the Pomeranchuk threshold $F_0 = -1$. The longitudinal mode ($\alpha=0$, Fig. \ref{fig_dens}a) is underdamped and the transverse mode ($\alpha = \pi/2$, Fig. \ref{fig_dens}c) is overdamped. For repulsive interactions, two ZS modes appear and disperse with increasing repulsion. Both are underdamped with a decay rate dictated by the disorder, as is evident from the abrupt change in spectral width upon crossing $s=1$, as the poles move above the particle-hole threshold. Interestingly, while in the longitudinal sector there appears to be one mode that just changes its character from damped to underdamped, in the transverse sector the two modes appear unconnected. This is evident from the fact that the spectral intensity decays as one goes from attraction to repulsion in Fig. \ref{fig_dens}c.

\begin{figure*}[t]
\includegraphics[width=\hsize]{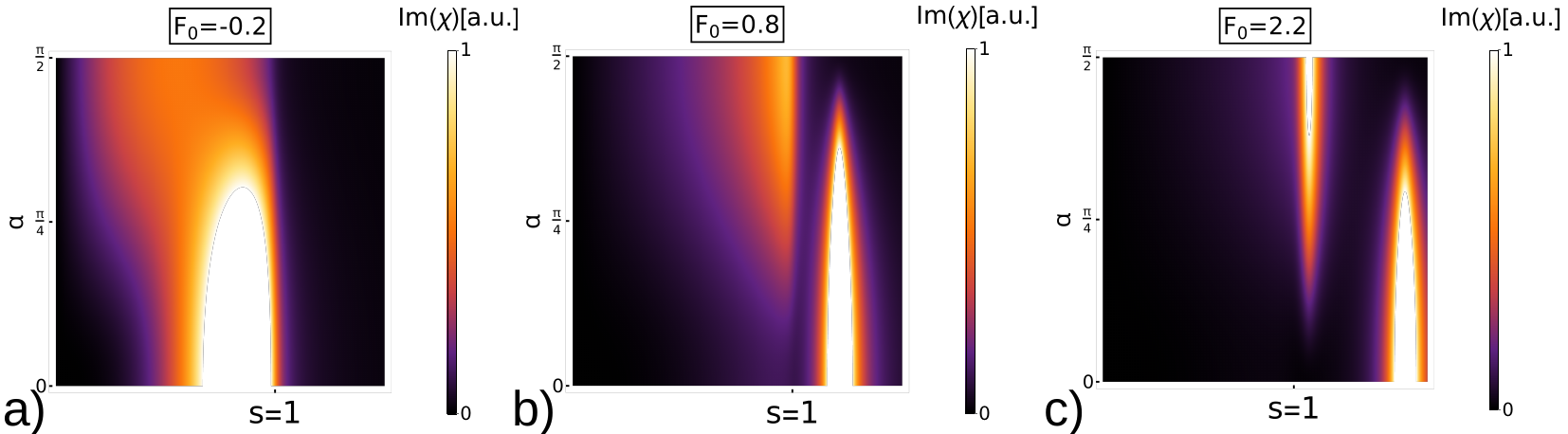}
\caption{\label{fig_pl3}  The angular dependence of the spectral function $\mbox{Im} \chi (\alpha, s)$, analogous to Fig. \ref{fig_dens}, is shown for three representative interaction strengths $F_0$.
%, with a fixed dimensionless disorder parameter $\delta$. The color scale represents the intensity of $\mbox{Im} \chi$ in arbitrary unit, with brighter regions indicating higher spectral weight. The frequency is fixed at$\frac{\w}{v_F q}$, illustrating the characteristic behavior of the spectral modes at this point. 
(a) Two damped modes, centered around $\alpha=0$ and $\alpha= \frac{\pi}{2}$. 
(b) A single mode appears dominant; however, this representation is somewhat deceptive as the transverse mode is hidden (see Sec.  \ref{sec3b}). (c) Two well-separated underdamped modes are evident. These are the mirage modes of the longitudinal and transverse sectors, and the transverse mode corresponds to the hidden mode seen in panel (b).}
%This progression indicates that increasing $F_0$ enhances mode separation and reduces damping effects. The choice of $\delta =.025$ ensures finite disorder-induced broadening while preserving distinct mode structures.}
\end{figure*}

\begin{figure*}[t]
\includegraphics[width=\hsize]{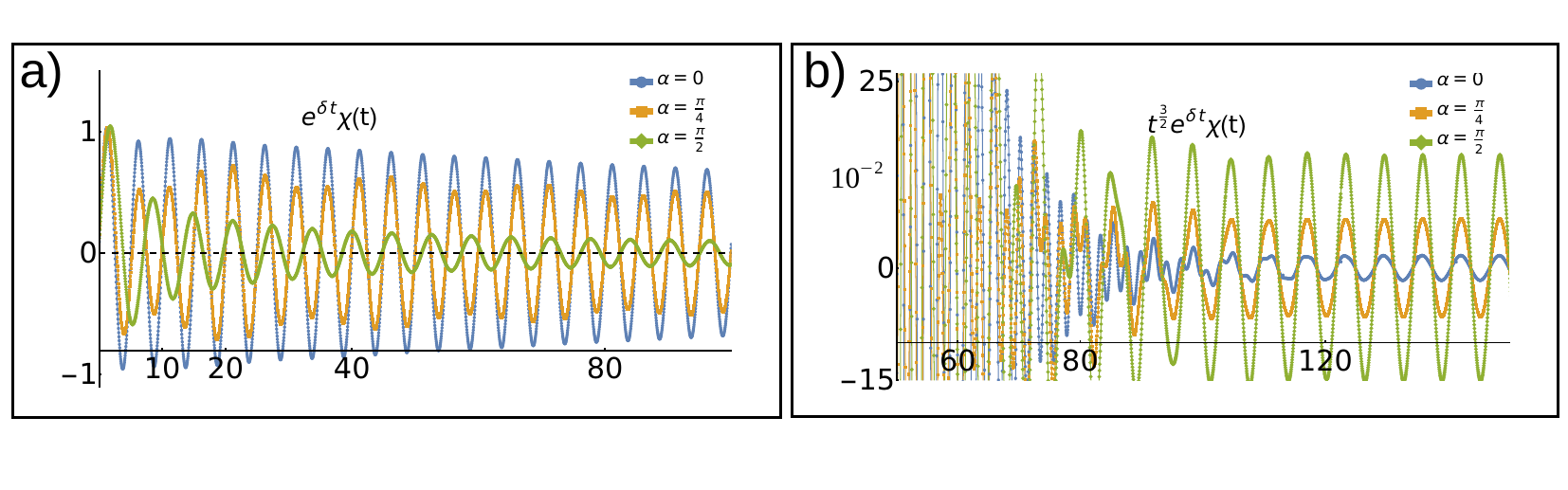}
\caption{\label{fig_pl1} Time evolution of hidden and mirage modes. (a) Short time behavior when the longitudinal mode is in the mirage regime and the transverse mode is hidden. As can be seen, the longitudinal mode is oscillating faster, and it can be verified that the trasverse mode is oscillating at precisely $s=1$, see Eq. \eqref{eq:chit-hidden-l0}. (b) Long time behavior in the regime with two mirage modes. As can be seen, both modes cross over to $s=1$. In both figures, we multiplied the numerical result by the analytic prediction for clarity. The numerical values here are respectively $F_0 = 0.95, 10$.}
\end{figure*}

Upon closer look, one may note that for repulsive interactions, not just the mode dispersions evolve with interaction but so does their width. The evolution of the width with interaction is a sign that these are so called ``mirage'' modes \cite{kleinhidden_9}, in the sense that they appear as well-formed Lorentzians in the spectral function (see Fig. \ref{fig_pl3}) but have a decay rate that is \emph{greater} than the disorder decay rate. Indeed, for a range of repulsive interaction strengths, the poles are located on the unphysical sheet below the branch cut, see Fig. \ref{fig_pl}b,d. The implication of this on the time dependent response is quite profound. The inset of Fig. \ref{fig_pictorial} depicts the time-dependent response for different $\alpha$ for two mirage modes. As can be seen, the decay time varies strongly with angle, creating the impression that the system is characterized by anisotropic scattering, e.g. polarization dependent disorder. In fact, this is not the case, and the way to verify this is by studying the long-time behavior. As can be seen in the Fig. \ref{fig_pl1}b, at long times the oscillation period changes from an \emph{angular dependent} ZS frequency ($\mbox{Re}s > 1$) to the frequency of the particle-hole threshold's branch point ($\mbox{Re}s\approx 1$), \emph{independent of angle}.

This behavior is in stark contrast to what would appear in a FL, e.g. in the spin sector. For an isotropic interaction, there is only one mode, which for an attractive interaction is either a damped mode, similar to the transverse mode in the attractive region, or a hidden mode (Fig. \ref{fig_pl}e). For a repulsive interaction the ZS frequency disperses with repulsion strength, but the width does not and the mode always resides on the physical sheet, above the branch cut, see Fig. \ref{fig_pl}f. Such hidden modes also appears in the Dirac liquid. Fig. (\ref{fig_pl3}) depicts the angular behavior of the spectral function for three representative interaction strengths. As can be seen, and as discussed above, in the transverse sector there appear to be two disconnected modes. Indeed, in Fig (\ref{fig_pl3})a one sees two damped modes, centered around $\alpha=0,\pi/2$, while in Fig. (\ref{fig_pl3})c one sees two clear underdamped modes. However, in Fig. (\ref{fig_pl3})b there is only one mode. Such a picture is of course an illusion, stemming from the fact that in the range $0< F_0 < F_{hm}$, the transverse pole is hidden. It is on the unphysical sheet just above the branch cut, a situation which is infinitesimally close to having the pole on the physical sheet but below the branch cut. The impact on the time-dependent response is the same: the spectral weight is concentrated at $s\lesssim 1$, but the actual mode is determined by the pole position, see Eq. \eqref{eq:chit-hidden-l0}. 
The time-dependent response for an interaction strength in this regime is depicted in Fig. \ref{fig_pl1}a. As can be seen there, the transverse and longitudinal components are oscillating at different frequencies. Indeed, it can be checked that the transverse component is oscillating at precisely $s=1$, and the longitudinal one at a frequency $s>1$.
The time-dependent response for the interaction strength in Fig. \ref{fig_pl3}c is in the inset of Fig. \ref{fig_pictorial}. However, the short time behavior for both regimes is an illusion, as can be seen by looking at Fig. \ref{fig_pl1}b.
%, using somewhat different parameters for clarity. 
At long times, all mirage modes eventually cross-over to the dominant behavior, which is just that of the branch point, $s=1$.

The mechanism leading to the qualitative change in the collective mode behavior is related to the topological number that is most associated with the local geometric structure of the Riemann surface, namely the number of poles. For a FL, the algebraic structure generated by the RPA sum is a second-order polynomial. The introduction of the form-factors associated with the chiral wavefunctions promotes this structure to a sixth-order polynomial, giving rise to a far richer behavior.

\subsection{\label{sec3c}The density-density response $\chi_{00}$}

Before ending, we briefly calculate the response in the density-density channel within our model. The calculation is somewhat irrelevant from a physical standpoint, as it neglects the long-range Coulomb repulsion, which dominates over the local interaction and introduces a plasmon mode. However, it serves to further clarify some of the algebraic properties.
so it is worth a brief discussion.

For $\chi_{00}$, the sum in Fig. \ref{fig_crosspol} has the form,
\begin{equation}
    \chi_{00} = \chi^b_{00} - \chi^b_{0j}\chi^b_{j0}\frac{F_0}{1+F_0 f_1 (s)},
\end{equation}

where $\chi^b_{0j}$ is a mixed polarization, with $\tau_j$ on one vertex and $\tau_0$ on the other. For the single cone considered in our work up to now, it is:
\begin{equation}
\label{eq:ddr}
\chi^b_{0j} = %\tau_j\cdot
\hq_j s l_0(s), \chi^b_{j0}=\chi^b_{0j}.
\end{equation}
Thus, the longitudinal mode shows up also in the density response. 

In practice, however, more care is needed, since a typical system hosts multiple cones. Consider for example the simplest case of two cones with opposite dispersions,
\begin{equation}
    \hat{H}_\pm = \sum_{\k} c^\dagger_{\k,\pm}\left(-\mu\tau_0 \pm v_F\bv{\tau}\cdot\k\right)c_{\k,\pm},     
\end{equation}
where $\pm$ denote the two cones. It is easy to see that the form factors in this case are,
\begin{equation}
     \bra{u_{\k}}\bv{\tau}\ket{u_{\k}}_\pm = \pm \left(\begin{array}{c}
          \cos\theta_{\k}\\\sin\theta_{\k}\\0
    \end{array}\right).
\end{equation}
The implication is that when summing up the two FS contributions, they exactly cancel out. It can be verified that the six cones in e.g. the tight-binding model for graphene also cancel out. Thus, unless the symmetry of the different FSs is broken somehow, we will not see the mode.

\section{Summary and Discussion}
\label{sec4}

In this work, we showed that the band geometry of an electron system has a significant impact on the ZS collective modes of the system, despite the fact that these modes involve excitations that are infinitesimally close to the electron FS. We found, that near a Dirac point the collective mode spectrum is qualitatively modified by the chiral nature of the wavefunctions, and that this shows up in the Riemann surface structure of the response functions encoding these excitations. In particular, for an isotropic (angular momentum $\ell=0$) interaction in the pseudospin sector, we found hidden and mirage modes (see description above), that are absent in the spectrum of a conventional FL. The existence of these modes, and their evolution with the Landau parameter $F_0$, which depends on e.g. the density of states, can serve as sensitive indirect probes for the existance of band-crossings in the Brillouin zone, when direct probes such as ARPES fail. In addition, as we showed above, a naive analysis of the spectral signature of these modes can be very misleading, and give the impression of anisotropic scattering mechanisms, or of independent collective modes, that are in fact nothing but the manifestation of the Riemann surface's structure, especially its branch points.

The main impact of the band geometry was to introduce an effective anisotropy in the polarization bubble via the chiral form factors. While anisotropic form factors are of course very common, the modification due to the band geometry is somewhat different from typical scenarios that introduce anisotropy in the particle-hole response. One common source of anisotropy is lattice effects, but these usually manifest as inversion-even anisotropies (even angular momenta) which break the continuous rotational symmetry to a discrete lattice one (e.g. \cite{Wegehaupt1978,Metzner2003,Devereaux2004}). \AK{
In addition, such anisotropies usually \emph{modify} the collective mode spectrum, e.g. by adding $\ell=2$ harmonics to the dominant $\ell=0$ scattering, assuming isotropic interactions. It is very hard to imagine a scenario where such lattice effects on the band-structure would be enough to change the behavior qualitatively to a different angular momentum channel. In Appendix \ref{appC}, we show this by studying two paradigmatic examples of a parabolic (rather than Dirac) band structure: Bernal-stacked bilayer graphene \cite{McCann_2013}   and the  two-orbital model for Fe-based superconductors \cite{Graser2010FeSC, PhysRevB.77.220503}. One finds that for the bilayer graphene case (with nonzero Berry connection), the collective mode spectrum is dominated by the $\ell = 2$ channel, with similar phenomenology to the Dirac case. In stark constrast, the two-orbital model, while having significant anisotropy and nontrivial form-factors (but zero Berry connection), is still predominantly $\ell=0$.
}

A second source of anisotropy is the appearance of anisotropic interactions, encoded in Landau parameters $F_{\ell > 0}$, e.g. nematic 
 or polar fluctuations \cite{Metzner2003,wu2007fermi_28,PhysRevResearch.1.033134_15}. The main difference between the effect of anisotropic interactions and band geometry is that the former can drive symmetry-breaking (Pomeranchuk) transitions, while the band effects do not. In addition, while we did not explicitly discuss it, our treatment can be readily generalized to the case of nonzero $F_{\ell >0}$, and the effect will be to mix different angular momentum channels together, giving rise to a far richer collective mode spectrum.

 We end by briefly touching on the topological aspects in this work. Broadly speaking, our system hosts three topological numbers: the Chern number for the fermionic bands \AK{(well-defined if we introduce an infinitesimal gap which doesn't affect the geometry)}, the genus of the polarization bubble Riemann surface, and order (in $s$) of the polynomial defining it. It appears from our work, that only the last of these is sensitive to the coupling of the fermionic and bosonic structures. This is probably because we did not consider symmetry-breaking fluctuations. These have the potential of splitting degenerate bands (which can cause a change of genus) and lifting topological protections (which can change the Chern number). Finally, we note that in this paper we only considered the zeroth order contributions in small $v_F|\q|/\mu$. Going to higher order can give rise to contributions arising from additional band-geometric quantities, such as the quantum metric \cite{Torma2022,Torma2023, Xiao2024QuantumGE_30, kang2025quantum_16}. Interestingly (see Appendix \ref{appA}) for the case of a Dirac cone these contributions, at least in the density-density channel, exactly cancel out with density of states corrections, but we expect them to appear in more complicated band structures.
 We leave such studies to further work.
 
\section*{Acknowledgments}
We thank A.V. Chubukov, R.M. Fernandes and J. Schmalian for helpful discussions. We acknowledge support by the Israel Science Foundation (ISF), and the Israeli Directorate for Defense Research and Development (DDR\&D) under grant No. 3467/21, and by the  United States - Israel Binational Science Foundation (BSF), grant No. 2022242.

\newpage
%\clearpage

\onecolumngrid
\appendix

%\onecolumn
%\section*{Appendix}

%\section{\label{app1} 
\section{\label{appA}Computation details of the bare particle-hole susceptibility (the polarization bubble)}

In this appendix, we provide detailed calculations for the polarization bubbles used in the main text.  
%We closely follow Ref. \cite{kleinhidden_9}, providing detailed explanations where necessary. 
Our primary focus is on the manner in which the chiral form-factors modify the functional dependence on $s$, so for simplicity in this appendix we perform the calculations in the clean limit. The generalization to static disorder is standard and follows e.g. Ref. \cite{PhysRevResearch.1.033134_15}.

As discussed in the main text, the general form of the (bare) polarization bubble is given in Eq. \eqref{eq:chi-bare-sublattice}, repeated here for clarity:
\begin{equation}
  \label{eq:chi-bare-sublattice-app}
  \chi_{\mu\nu}^{b}(\q,i\W) = -\frac{1}{2}\int \frac{d\w d^2k}{\tpp^3} \mbox{Tr}\left[ \tau_{\mu} \hat{G}(\k_+,\w_+)\tau_{\nu} \hat{G}(\k_-,\w_-)\right],
\end{equation}
where  $\omega_\pm = \omega \pm \frac{\W}{2}$ and $\k_\pm = \k \pm \frac{\mathbf{q}}{2}$, see Eq. \eqref{eq:Pi-def-intro}, and $\mu,\nu=0,x,y$ (the bubbles with a $z$ are zero for a Dirac cone in 2D).

In this Appendix we rescale all momenta $\k,\q\to v_F\k,v_F\q$ in order to reduce the algebra somewhat. The main text includes the expressions with the $v_F$ units returned.

%only the upper band intraband contribution is relevant
\subsection{\label{appA1} Bare polarization bubble with $\tau_0$ on both vertices}

We start with the retarded density-density response function, $\chi^b_{00}$. Summing over the sublattice degrees of freedom, we find,

\begin{align}
%\label{eq:Pi-tot}
\chi^b_{00} (\q, i \W)
%&= T\sum_n \int \frac{d^2k}{\tpp^2}  \mbox{Tr}\left[ \tau_0 \hat{G}(\k_+,i\w_+)\tau_0 \hat{G}(\k_-,i\w_-)\right] \nn\\
&= \int\frac{d\w d^2k}{\tpp^3}  \frac{i(\w_--i\mu)i(\w_+-i\mu) + \k_-\cdot\k_+}{\left((\w_- - i \mu)^2 + |\k_-|^2\right)\left((\w_+ - i \mu)^2 + |\k_+|^2\right)} \nn\\
&=  \int\frac{d\w d^2k}{\tpp^3}  \left[i(\w_--i\mu)i(\w_+-i\mu) + 
%|\k_-||\k_+|\cos\Delta\theta
\k_-\cdot\k_+
\right]G^+_-G^+_+G^-_-G^-_+,\label{eq:Pi-diag-2}
\end{align}

where $\Delta\theta$ is the angle between $\k_-,\k_+$. In Eq. \eqref{eq:Pi-diag-2} we have defined the following Green's functions, such that the upper index denotes particle or hole, and the bottom index denotes sum or difference,
\begin{equation}
  \label{eq:G-def}
  G^\alpha_\beta = \left(i(\w_\beta-i\mu) - \alpha |\k_\beta|\right)^{-1}.
\end{equation}

Now we note the following simple relations,
\begin{align}
  \label{eq:G-rels}
  i(\w_\beta-i\mu) &= (G_\beta^\alpha)^{-1}+\alpha |\k_\beta|, \\
  |\k_\beta| &= -\alpha(G_\beta^\alpha)^{-1} + \alpha i(\w_\beta-i\mu), \label{eq:G-rels-1} \\
  G^-_\beta G^+_\beta &= \left(G^-_\beta+G^+_\beta\right)\frac{1}{2i(\w_\beta-i\mu)}=\left(G^-_\beta-G^+_\beta\right)\frac{\alpha}{2|\k_\beta|}\, . \label{eq:G-rels-2}
\end{align}
We integrate over $\omega$ first.
%$\chi^b_{00} (\q, i \W)$ over the poles $\w$ in a brute-force manner, starting from Eq. \eqref{eq:Pi-diag-2}. 
The Green's functions, written explicitly, are:
\begin{equation}
  \label{eq:Green-list}
  \left[\left(i\w_-+(\mu-|\k_-|)\right)\left(i\w_- +(\mu+|\k_-|)\right)\left(i\w_+ +(\mu - |\k_+|)\right)\left(i\w_+ +(\mu+|\k_+|)\right)\right]^{-1}.
\end{equation}
Clearly, the hole propagators with $\mu + |\k_\pm|$ are always in the upper half-plane. So, we integrate over the lower half plane and obtain,
\begin{flalign}
  \label{eq:Pi0-poles-1}
  \chi^b_{00} (\q, i \W) =- \int \frac{d^2k}{\tpp^2} \left\{
  \Theta(|\k_-|-\mu)\frac{|\k_-|(|\k_-|+i\W) + \k_-\cdot\k_+}{2|\k_-|(|\k_-|-|\k_+|+i\W)(|\k_-|+|\k_+|+i\W)}\right.\nn\\
  +\left.
  \Theta(|\k_+|-\mu)\frac{|\k_+|(|\k_+|-i\W) + \k_-\cdot\k_+}{2|\k_+|(|\k_+|-|\k_-|-i\W)(|\k_+|+|\k_-|-i\W)}
  \right\}\, .
\end{flalign}
Now, notice that the denominator products in the expression have exactly the same form as the Green's functions products in Eq. \eqref{eq:G-rels-2}. This means we can again split them into either sums or differences, e.g.
\begin{align}
  \label{eq:G-rels-3}
  \frac{1}{(|\k_-|-|\k_+|+i\W)(|\k_-|+|\k_+|+i\W)} &= \frac{1}{2(|\k_-| + i\W)}\left[\frac{1}{|\k_-|-|\k_+|+i\W}+\frac{1}{|\k_-|+|\k_+|+i\W}\right]\nn\\
  &= \frac{1}{2|\k_+|}\left[\frac{1}{|\k_-|-|\k_+|+i\W}-\frac{1}{|\k_-|+|\k_+|+i\W}\right].
\end{align}

So, splitting the propagators, adding them up for the $|\k_{\pm}|(|\k_{\pm}| \pm i\W)$ terms and subtracting for the $\k_- \cdot \k_+$ terms, we finally end up with,

\begin{align}
  \label{eq:Pi0-poles-2}
  \chi^b_{00} = 
  \frac{1}{4} 
  \int\frac{d^2k}{\tpp^2} (T_{intra} + T_{inter}).
\end{align}
where the intraband contribution is,
\begin{align}
  \label{eq:T1-def}
  T_{intra} &= \frac{1+\cos\Delta\theta}{2}\frac{\Theta(|\k_-|-\mu)-\Theta(|\k_+|-\mu)}{|\k_-|-|\k_+|+i\W} \nn\\
      &= \frac{1+\cos\Delta\theta}{2}\frac{\Theta(\mu-|\k_+|)-\Theta(\mu-|\k_-|)}{|\k_-|-|\k_+|+i\W}.
\end{align}
Here $\Delta\theta$ is angle between $\k_+$ and $\k_-$, and we flip the Heaviside functions to bring the expression to the ``standard'' Fermi function forms. The angular factor $(1/2)(1+\cos\Delta\theta)$ is precisely the chiral form-factor from Eq. \eqref{eq:Pi-def-intro}.

The interband term is,
\begin{align}
  \label{eq:T1-def-2}
  T_{inter} &= \frac{1-\cos\Delta\theta}{2}\left[\frac{\Theta(|\k_-|-\mu)}{|\k_-|+|\k_+|+i\W} + \frac{\Theta(|\k_+|-\mu)}{|\k_-|+|\k_+|-i\W}\right],
\end{align}
where again the angular term is the form-factor, this time between bands. It is easy to see that for finite $\mu$, $T_{intra}$ has a singularity for small $\q,\W$, and is $O(1)$ whereas $T_{inter}$ is smaller by order $\W/\mu$. On the other hand for $\mu=0$ we have
\begin{align}
  \label{eq:T1-T2-neutrality}
  T_{intra} = 0, T_{inter} = \frac{2(1-\cos\Delta\theta)(|\k_-|+|\k_+|)}{(|\k_-|+|\k_+|)^2+\W^2}.
\end{align}

In what follows, we evaluate 
%Now, let's evaluate 
$T_{intra}$ on the Matsubara axis. We go beyond the leading order found in the main text, trying to expand in small $|\q| \ll \mu$. The reason can be seen by expanding the form-factor,
\begin{equation}
    \bra {u_{\k_-}} \tau_{0} \ket {u_{\k_+}}  \bra {u_{\k_+}} \tau_{0} \ket {u_{\k_-}} =  \left|\braket{u_{_{\k_-}}|u_{_{\k_+}}}\right|^2 
    = 1 - q_i g_{ij} q_j + \cdots,
\end{equation}
where
\begin{equation}
    g_{ij} = \braket{\pd_{\k_i}u_{\k}|\pd_{\k_j}u_{\k}}-\braket{\pd_{\k_i}u_{\k}|u_{\k}}\braket{u_{\k}|\pd_{\k_j}u_{\k}},
\end{equation}
which is the quantum metric, i.e. the Harmonic conjugate of the Berry curvature. The various terms are:
\begin{align}
  \label{eq:T1-terms}
  \cos \Delta\theta &= \frac{(\k+\q/2)\cdot(\k-\q/2)}{|\k+\q/2||\k-\q/2|} = \frac{k^2-q^2/4}{\sqrt{(k^2+k q \cos\theta+q^2/4)(k^2-k q \cos\theta+q^2/4)}} \nn\\
                    &= 1 - \frac{q^2}{2k^2}\sin^2\theta + \cdots \nn \\
  (|\k_-|-|\k_+|+i\W)^{-1} &= \frac{1}{q (\cos\theta - i s)} \left[1 + \frac{q^2}{8k^2(\cos\theta-i s)}\cos\theta\sin^2\theta+\cdots\right] \nn \\
  \Theta(\mu - |\k_\pm|) &= \Theta(\mu^2 -(k^2\pm k q \cos\theta+q^2/4) = \Theta\left(\sqrt{\mu^2 - q^2\sin^2\theta/4} \mp q \cos\theta /2 - k\right) \nn\\
                    &= \Theta\left(\mu \mp q\cos\theta/2 - \frac{q^2}{8\mu^2}\sin^2\theta + \cdots - k \right).
\end{align}

Now, we can perform the $k$ integration,

\begin{align}
  \label{eq:T1-expand-integrate}
 \chi^b_{00}(\theta)  &=  \frac{1}{2}\int_0^\infty k dk T_{intra} \nn\\
                 &= \mbox{sign}(\cos\theta)\int_{\mu_<}^{\mu_>} \frac{k dk}{q (\cos\theta - is)}\left[\left(1 - \frac{q^2}{4k^2}\sin^2\theta\right)\left(1 + \frac{q^2}{8k^2(\cos\theta-i s)}\cos\theta\sin^2\theta\right) +\cdots\right] \nn\\
                 &= \mbox{sign}(\cos\theta)\int_{\mu_<}^{\mu_>}\frac{k dk}{q (\cos\theta - is)}\left(1 - \frac{q^2}{4k^2}\sin^2\theta  + \frac{q^2}{8k^2(\cos\theta-i s)}\cos\theta\sin^2\theta + \cdots\right) \nn\\
                 &= \mbox{sign}(\cos\theta)\left[ \frac{\mu_>^2-\mu_<^2}{2q(\cos\theta-i s)}+\frac{\log\left(\frac{\mu_>}{\mu_<}\right)}{q\cos\theta - i s}\left(- \frac{q^2}{4}\sin^2\theta + \frac{q^2}{8(\cos\theta-i s)}\cos\theta\sin^2\theta \right)+\cdots\right] \nn\\
                 &= \frac{\mu\cos\theta}{\cos\theta-i s}\left(1-\frac{q^2}{8\mu^2}\sin^2\theta - \frac{q^2}{4\mu^2}\sin^2\theta+\frac{q^2}{8\mu^2(\cos\theta-i s)}\cos\theta\sin^2\theta + \cdots\right).
\end{align}
Here, the first $q^2$ contribution comes from the band structure, the second from the form factor, and the third from the energy difference. For convenience we defined $\mu_{<>} = \mu \pm q|\cos\theta|/2 - q^2\sin^2\theta/(8\mu^2)$. Crucially, \emph{all three terms} have contributions proportional to $s^3$. Indeed, performing the integrals we find,

\begin{align}
  \label{eq:angular-ints}
   \int \frac{d\theta}{2\pi} \frac{\cos\theta}{\cos\theta - is} &= l_0(i s)\\
  \int \frac{d\theta}{2\pi} \frac{\cos\theta\sin^2\theta}{\cos\theta - is} &= \int \frac{d\theta}{2\pi} \frac{\cos\theta-\cos^3\theta}{\cos\theta - is} \\
  &= l_0(i s) - \frac{1}{2}l_1(i s)\\
   \int \frac{d\theta}{2\pi} \frac{\cos^2\theta\sin^2\theta}{(\cos\theta - is)^2} &= \frac{d}{d(is)}\int \frac{d\theta}{2\pi} \frac{\cos^2\theta-\cos^4\theta}{\cos\theta - is} = \frac{d}{d(is)}\left[i s \left(l_0(i s) - \frac{1}{2}l_1(i s)\right)\right] \nn\\
   &= 2 l_0(i s) - \frac{3}{2}l_1(i s) \nn
\end{align}

Thus, the final result is,
\begin{equation}
  \label{eq:Pi1-final}
  \chi^b_{00} = \mu l_0(is) \left(1 - \frac{q^2}{8\mu^2}\right) + \cdots
\end{equation}
and the band structure + energy factor exactly cancel out the $l_1$ from the form factor (i.e. the contribution from the quantum metric). If we only consider zeroth order of $q$, this result gives Eq. \eqref{eq:Pi-final} of the main text.

\subsection{\label{appA2} The polarization bubble with $\tau_{0}$ and $\tau_{j}$ in the vertices}

In this section, we calculate the cross-polarization, following the derivation in Sec. \ref{appA1}. 
%Specifically, we compute the susceptibility with a $\tau_0$ at the right-hand vertex while keeping the left-hand vertex as $\tau_{j}$, where $j=x,y$ (see bare bubble in Fig. \ref{fig_crosspol}). 
%Following the same procedure as before, w
We write,
\begin{align}
\label{eq:Pi-sublat-full1}
    \chi^b_{0j} (\q, i \W)&=  \frac{1}{2}\mbox{Tr} \int\frac{d\w d^2k}{\tpp^3}\left(i (\w_- - i \mu)\tau_0 - \tau_m(\k_-)_m\right)\tau_0\left(i (\w_+ - i \mu)\tau_0 - \tau_n(\k_+)_n\right)\tau_j G^+_-G^+_+G^-_-G^-_+\nn \\
    %\int\frac{d\w d^2k}{\tpp^3} \mbox{Tr}\left[ \hat{G}(\k_+,i\w_+)\tau_{i}\hat{G}(\k_-,i\w_-)\tau_{j}\right]\, , 
    &= \int\frac{d\w d^2k}{\tpp^3}\left[-i (\w_- - i \mu) (\k_+)_n \delta_{nj} -i (\w_+ - i \mu) (\k_-)_m \delta_{mj}+ i (\k_-)_m (\k_+)_n \epsilon_{mnj}\right] G^+_-G^+_+G^-_-G^-_+\nn \\
    &=\int\frac{d\w d^2k}{\tpp^3}\left[-i (\w_- - i \mu) (\k_+)_j -i (\w_+ - i \mu) (\k_-)_j +i (\k_-\times\k_+)_j\right] G^+_-G^+_+G^-_-G^-_+, 
\end{align}
%where repeated indices $m,n$ are summed over but repeat $j$ indices are not. 
Performing the Matsubara frequency summation leads to, (similar to \eqref{eq:Pi0-poles-1}),
\begin{align}
  \label{eq:Pi-AB-1}
\chi^b_{0j} (\q,i \Omega) %&= \int \frac{d^2k}{\tpp^2} \left\{\Theta(|\k_-|-\mu)i\frac{(-i|\k_-|) \tau_j (\k_+)_j + (\W-i|\k_-|)\tau_j (\k_-)_j}%(\W/2-i|\k_-|)\bvs\cdot\k - \W\bvs\cdot \q}
             %      {(-2i|\k_-|)(\W-i|\k_-|+i|\k_+|)(\W-%i|\k_-|-i|\k_+|))}\right.\nn\\
                 %&\qquad\qquad\qquad\left.+\Theta(|\k_+|-\mu)i\frac{(-\W-i|\k_+|)\tau_j (\k_+)_j + (-i|\k_+|)\tau_j (\k_-)_j}
                   % (-\W/2-i|\k_+|)\bvs\cdot\k - \W\bvs\cdot\q}
                   %{(-2i|\k_+)|(-\W-i|\k_+|+i|\k_-|)(-\W-i|\k_+|-i|\k_-|)}%\right\} \nn\\
                 &= \int \frac{d^2k}{\tpp^2}  \left\{\Theta(|\k_-|-\mu)\frac{(|\k_-|)(\k_+)_j  +(|\k_-|+i\W) (\k_-)_j+i (\k_-\times\k_+)_j}
                   % (\W/2-i|\k_-|)\bvs\cdot\k - \W\bvs\cdot \q}
                   {(2|\k_-|)(|\k_-|-|\k_+|+i\W)(|\k_-|+|\k_+|+i\W))}\right.\nn\\
                 &\qquad\qquad\qquad\left.+\Theta(|\k_+|-\mu)\frac{(|\k_+|-i\W) (\k_+)_j + (|\k_+|) (\k_-)_j+i (\k_-\times\k_+)_j}
                   % (-\W/2-i|\k_+|)\bvs\cdot\k - \W\bvs\cdot\q}
                   {(2|\k_+|)(|\k_+|-|\k_-|-i\W)(|\k_+|+|\k_-|-i\W)}\right\}. 
\end{align}
So, similarly to $ \chi^b_{00}$, we can write by using Eq. \eqref{eq:G-rels-3}
\begin{align}
  \label{eq:Pi-AB-2}
  \chi^b_{0,j} (\q,i \Omega) = \frac{1}{4}\int \frac{d^2k}{\tpp^2} (U_{intra,j} + U_{inter,j}),
\end{align}
where,
\begin{align}
  \label{eq:U1-def}
  U_{intra,j} &=  \frac{\Theta(|\k_-|-\mu)-\Theta(|\k_+|-\mu)}{i\W + |\k_-| - |\k_+|}  \left((\hk_-)_j+(\hk_+)_j+i (\hk_-\times\hk_+)_j\right),
\end{align}
and
\begin{align}
  \label{eq:U2-def}
  \hat U_{inter} &= \left[\frac{\Theta(|\k_-|-\mu)}{i\W + |\k_-| + |\k_+|} - \frac{\Theta(|\k_+|-\mu)}{-i\W + |\k_-| + |\k_+|}\right] \left((\hk_-)_j-(\hk_+)_j-i (\hk_-\times\hk_+)_j\right).
\end{align}
Interestingly, $\hat U_{intra}$ has a nonzero contribution at zeroth order expansion in $q$. Indeed, at the lowest order we obtain
\begin{align}
  \label{eq:U1-leading}
   U_{intra} \approx  2 \frac{\Theta(\mu - q\cos\theta/2 - k)-\Theta(\mu + q\cos\theta/2 - k)}{i\W - q\cos\theta }\hat{k}_j,
\end{align}

While $\hat U_{inter} = 0$ to the same order, we approximate $k ~dk \approx \mu dk$. Changing variables to $\theta = \theta_k - \theta_q$ and integrating \eqref{eq:Pi-AB-2} over $k$, we obtain:

\begin{align}
  \label{eq:Pi-AB-3}
  \chi^b_{0j} (\q,i s) &\approx  \frac{\mu}{4\pi}\int d\theta \frac{q\cos(\theta_k-\theta_q)}{q\cos(\theta_k-\theta_q) - i\W } \begin{pmatrix}\cos\theta_k \\ \sin\theta_k\end{pmatrix}_j \nn\\
  & = \frac{\mu}{4\pi}\int d\theta \frac{\cos\theta}{\cos\theta-i s} \begin{pmatrix} \cos\theta \cos\theta_q - \sin\theta_q \sin\theta\\\sin\theta \cos\theta_q+\cos\theta\sin\theta_q \end{pmatrix}_j \nn \\
  & =  \frac{\mu}{4\pi}\int d\theta \frac{\cos^2\theta}{\cos\theta-i s}\begin{pmatrix}\cos\theta_q \\ \sin\theta_q\end{pmatrix}_j \nn \\
  %& =  \frac{\mu}{4\pi}  2 i \pi |s| (1- \frac{|s|}{\sqrt{1+s^2}}) \hq_j \nn \\
  %&= \frac{i \mu}{2} \hq_j|s|\left(1- \frac{|s|}{\sqrt{1+s^2}}\right)\\
  & = \frac{i \mu}{2} |s| l_0(is) \hq_j
\end{align}
where we used the fact that odd integrations over $\theta$ vanish, and that  $\hq = (\cos\theta_q,\sin\theta_q)$. Extending to the real axis then yields Eq. \eqref{eq:ddr}.

\subsection{\label{appA3} The polarization bubble with $\tau_{i}$ and $\tau_{j}$ in the vertices}

Following the same procedure as before, we write,
\begin{align}
\label{eq:Pi-sublat-full}
    \chi^b_{ij} (\q, i \W)&=  \frac{1}{2}\mbox{Tr} \int\frac{d\w d^2k}{\tpp^3}\left(i (\w_- - i \mu)\tau_0 - \tau_m(k_-)_m\right)\tau_i\left(i (\w_+ - i \mu)\tau_0 - \tau_n(k_+)_n\right)\tau_j G^+_-G^+_+G^-_-G^-_+\nn\\
    %\int\frac{d\w d^2k}{\tpp^3} \mbox{Tr}\left[ \hat{G}(\k_+,i\w_+)\tau_{i}\hat{G}(\k_-,i\w_-)\tau_{j}\right]\, , 
    &= \int\frac{d\w d^2k}{\tpp^3}\left[i (\w_- - i \mu)i (\w_+ - i \mu) - \k_-\cdot\k_+]\delta_{ij} +[(k_-)_i(k_+)_j+(k_-)_j(k_+)_i\right] G^+_-G^+_+G^-_-G^-_+,
\end{align}
 
Performing the Matsubara frequency summation leads to,
%the integral representation,
\begin{align}
  \label{eq:Pi-ij-2}
 \chi^b_{ij} (\q, i \W) %&= \int \frac{d^2k}{\tpp^2} \left\{\Theta(|\k_-|-\mu)i\frac{\left[i (-i|\k_-|)i (\W-i|\k_-|)-\k_-\cdot\k_+\right]\delta_{ij} +(k_-)_i(k_+)_j+(k_-)_j(k_+)_i }
                 %  {(-2i|\k_-|)(\W-i|\k_-|+i|\k_+|)(\W-i|\k_-|-i|\k_+|))}\right.\nn\\
                % &\qquad\qquad\left.+\Theta(|\k_+|-\mu)i\frac{[i (-\W-i|\k_+|)i (-i|\k_+|) - \k_-\cdot\k_+]\delta_{ij} +[(k_-)_i(k_+)_j+(k_-)_j(k_+)_i]}
                 %  {(-2i|\k_+|)(-\W-i|\k_+|+i|\k_-|)(-\W-i|\k_+|-i|\k_-|)}\right\} \nn\\
                   &=  -\int \frac{d^2k}{\tpp^2} \left\{\Theta(|\k_-|-\mu)\frac{\left[|\k_-| (i\W+|\k_-|)-\k_-\cdot\k_+\right]\delta_{ij} +(k_-)_i(k_+)_j+(k_-)_j(k_+)_i }
                   {(2|\k_-|)(|\k_-|-|\k_+|+i\W)(|\k_-|+|\k_+|+i\W))}\right.\nn\\
                   &\qquad\qquad\left.+\Theta(|\k_+|-\mu)\frac{[(|\k_+|-i\W) |\k_+| - \k_-\cdot\k_+]\delta_{ij} +[(k_-)_i(k_+)_j+(k_-)_j(k_+)_i]}
                   {(2|\k_+|)(|\k_+|-|\k_-|-i\W)(|\k_+|+|\k_-|-i\W)}\right\} \nn\\
                   &= \frac{1}{4}\int \frac{d^2k}{\tpp^2}(V_{intra,ij}+V_{inter,ij}).
\end{align}
%Rewriting in terms of interband contributions, we obtain $V_1$:
We are interested only in the intraband term,
\begin{align}
    V_{intra,ij} &= \left[(1-\hat{k}_-\cdot\hat{k}_+)\delta_{ij}+(\hat{k}_-)_{i}(\hat{k}_+)_{j}+(\hat{k}_-)_{j}(\hat{k}_+)_{i}\right]\frac{\Theta(\mu-|\k_-|)-\Theta(\mu-|\k_+|)}{|\k_+|-|\k_-|-i\W} \\
    &\approx 2\hat{k}_{i}\hat{k}_{j}\frac{\Theta(\mu-|\k_-|)-\Theta(\mu-|\k_+|)}{|\k_+|-|\k_-|-i\W}
\end{align}
%
%Now after integrating with respect to $k$, 
%
%Carrying out the remaining integrations,
Thus,
\begin{equation}
\chi^b_{ij}  (\q, i \W)= \frac{1}{2} \int \frac{d^2 k}{(2\pi)^2} \hat{k}_{i} \hat{k}_{j} \frac{\Theta(\mu - \lvert \k_- \rvert) - \Theta(\mu - \lvert \k_+ \rvert)}{\lvert \k_+ \rvert + \lvert \k_- \rvert - i \Omega}
\end{equation}
This is (up to a global prefactor) the same bare bubble one would find for a single, nonzero Landau parameter $F_{\ell=1}$. Explicitly, it yields

\begin{align}
	\chi^b_{ij} (\q,s) &=\frac{\mu}{4\pi}
	\begin{bmatrix}
		 [(l_0-\frac{l_1}{2})+ \cos^2\theta_q (l_1-l_0)] &
		 (l_1 - l_0) \sin\theta_q \cos\theta_q\\
		(l_1 - l_0) \sin\theta_q \cos\theta_q &
		( l_0 -\frac{l_1}{2})+ \sin^2\theta_q (l_1 - l_0)
	\end{bmatrix}_{ij} \nn\\
    &= \frac{\mu}{8\pi} [(2l_0 - l_1) \delta_{ij}  + (2l_1-2l_0) \hq_{i} \hq_{j}] \nn \\
    &= \frac{\mu}{8\pi} [f_1(s) \hq \hq +f_2(s) \hat{q_t} \hat{q_t}   ]
\end{align}

which matches the  result given in Eq. \eqref{eq:anglechi}.

\subsection{\label{appA4} Computation of $\chi_{00}^b$ at the charge neutrality point}

Finally, for completeness, we rederive the well-known expression \cite{Wunsch_2006_11, hwang2007dielectric_31} for the polarization bubble at the charge neutrality point (CNP). The relevant expressions are given by
%, we have previously established that the term in 
Eqs.~\eqref{eq:T1-def-2} and
%dominates in Eq.~
\eqref{eq:Pi0-poles-2}. We now proceed with a detailed computation.

It is more convenient to perform the computation directly on the real axis. Upon analytically continuing to real frequencies, $i\Omega \to \Omega + i\delta$, at $\mu = 0$, we obtain:
\begin{align}
  \label{eq:T1-def1}
  T_{inter} &= \frac{1 - \cos\Delta\theta}{2} \left[\frac{1}{|\mathbf{k}_-| + |\mathbf{k}_+| + \Omega + i\delta} + \frac{1}{|\mathbf{k}_-| + |\mathbf{k}_+| - \Omega - i\delta}\right].
\end{align}
The polarization function can thus be expressed as:
\begin{align}
  \label{eq:Pi0-poles-3}
  \chi^b_{00} =  \int \frac{d^2k}{(2\pi)^2} T_{inter},
\end{align}

where the imaginary component of $T_{inter}$ is (for $\W > 0$),
%Focusing first on the imaginary part, integrating over $\theta$, we obtain:
\begin{equation}
    \text{Im}T_{inter} =
    %\,\chi^b_{00} = \int d\theta  
    \pi \left(1 - \frac{\mathbf{k}_- \cdot \mathbf{k}_+}{|\mathbf{k}_-||\mathbf{k}_+|}\right) \delta \left[\Omega - (|\mathbf{k}_-| + |\mathbf{k}_+|)\right].
\end{equation}
This expression is nonzero only for 
$\W > |\q|$. Now,
%utilizing the symmetry to $\k \to -\k$, and 
shifting the momentum $\mathbf{k} \to \mathbf{k} + \frac{\mathbf{q}}{2}$, we rewrite:
\begin{equation}
    \text{Im}\,\chi^b_{00} = \Theta(\W-|\q|)\int \frac{k dk}{8\pi}\int d\theta \left(1 - \frac{k + |\q|\cos\theta}{\sqrt{k^2 + |\q|^2 + 2k|\q| \cos\theta}}\right) \delta \left[\W - \left(k + \sqrt{k^2 + |\q|^2 + 2k|\q| \cos\theta}\right)\right],
\end{equation}
where $\Theta(x)$ is a Heaviside function.
Using the delta function property:
\begin{equation}
 \delta[g(\cos\theta)] = \frac{\delta(\cos\theta - x_0)}{|g'(x_0)|}
\end{equation}
with
\begin{equation}
 g(\cos\theta) = \Omega - \left(k + \sqrt{k^2 + |\q|^2 + 2k|\q| \cos\theta}\right), \qquad x_0 = \frac{(\Omega - k)^2 - k^2 - |\q|^2}{2 k |\q|}, % = 0,
\end{equation}
and%its derivative evaluates to:
\begin{equation}
 %g'(\cos\theta) 
 g'(x_0)= -\frac{k|\q|}{\sqrt{k^2 + |\q|^2 + 2k|\q| x_0}} = - \frac{k |\q|}{|\W-k|},
\end{equation}
we arrive at,
\begin{align}
    \text{Im}\chi_{00}^b &= \frac{\Theta(\W-|\q|)}{8\pi}\int_0^\infty k dk\int_{-1}^1 \frac{dy}{\sqrt{1-y^2}}  \left(1 - \frac{k+ |\q| y}{\sqrt{k^2 + |\q|^2 + 2k |\q| y}}\right)\frac{\delta(y-x_0)}{|g'(x_0)|} \\
    &= -\frac{\Theta(\W-|\q|)}{8\pi}\int_{\frac{\W-|\q|}{2}}^{\frac{\W+|\q|}{2}}dk\sqrt{\frac{|\q|^2-(\W-2k)^2}{\W^2-|\q|^2}}\\
    &= - \frac{|\q|}{32}\frac{\Theta(s-1)}{\sqrt{s^2-1}}.
\end{align}
Extending the function yields,
\begin{equation}
    \chi_{00}^b(s) = 
    %- \frac{|\q|}{32}\frac{1}{\sqrt{s^2-1}}.
    \frac{|\q|}{32}\frac{1}{\sqrt{1-s^2}},
\end{equation}
which is the same result as Eq. \eqref{eq:cnp}.

\section{\label{appB} Determining  $F^{hm}_0$ from Eq. \eqref{eq:F0-mh}}

$F_0^{hm}$ is defined as the interaction value for which the pole moves from being hidden to being mirage, i.e. when its imaginary part is exactly $-i\delta$.

The appropriate form of $f_2(s)$ is given by
\begin{equation} 
f_2(s) = 1 - 2s \left(s+i\delta 
%\mp 
+ i \sqrt{1-(s+i\delta)^2}\right),
\end{equation}
which is just the extension of Eq. \eqref{eq:f2-branch} to the unphysical sheet in the region $|s|<1$. 

To obtain the solution, we replace $s \rightarrow s'-i\delta$ and solve the real and imaginary parts of the pole equation for $F_0^{hm}$ and $s'$:

\begin{align}
 \label{eq:a1}
   \text{Re}\,\left(1+ F_0^{hm} f_2(s)\right)= \frac{1+F_0^{hm}}{F_0^{hm}} - 2 (s'-i\delta) (s'+i \sqrt{1-s'^2}) - 2 (s'+i\delta) (s'-i \sqrt{1-s'^2}) &= 0 \\
 \text{Im}\,\left(1+ F_0^{hm} f_2(s)\right)= -2 (s'-i\delta) (s'+i \sqrt{1-s'^2}) + 2 (s'+i\delta) (s'-i \sqrt{1-s'^2}) &= 0.\label{eq:a2}
\end{align}
The solution 
%Solving, \eqref{eq:a1} and \eqref{eq:a2} 
yields the result in main text, Eq. \eqref{eq:F0-mh}:
\begin{equation}
    F^{h,m}_0= 1; s= \pm \sqrt{1-\delta^2}.
\end{equation}

%\color{red}
\section{\label{appC} Parabolic dispersions: Berry non-trivial vs. trivial }

In the main text we focused on linear Dirac bands.  To demonstrate
that our collective–mode geometric effects extends to \emph{parabolic}
bands that carry finite Fermi-surface Berry phase,
we now work out two concrete examples.  Throughout this appendix we
use the same notation as in Secs. \ref{sec2}, \ref{sec3}.

\subsection{\label{appC1} Bernal-stacked bilayer graphene}

A canonical example of a parabolic band with nontrivial topology is that of Bernal-stacked bilayer graphene. The full tight-binding model is a 4-orbital one (two orbitals $A_i,B_i$ per layer, $i=1,2$). However, near the $K$ point the low-energy sub-space is spanned by the
\emph{non-dimer} orbitals $(A_1,B_2)$.  Projecting the four-band
Hamiltonian onto this sub-space gives the effective two-band model
\cite{McCann_2013}  
\begin{equation}
\mathcal{H}_{\text{eff}}(\mathbf{k})=
-\frac{1}{2m^*}
\begin{pmatrix}
0 & (\pi^\dagger)^{\!2}\\[4pt]
\pi^{2} & 0
\end{pmatrix}-\mu \tau_0,\qquad
\pi=k_x+i k_y,\quad
m^*=\frac{\gamma_1}{2v_F^{\,2}}.
\label{eq:Hbilayer}
\end{equation}
Here $\gamma_1$ is the inter-layer dimer hopping and
$v_F$ is the monolayer Fermi velocity. Diagonalizing Eq.~\eqref{eq:Hbilayer} yields the parabolic spectrum
\begin{equation}
E_\pm(\mathbf k)=\pm\frac{|\k|^2}{2m^*}-\mu.
\label{eq:Ebilayer}
\end{equation}
%so the Fermi contour is a circle of radius
%$q_F=\sqrt{2m^*|\mu|}/\hbar$.
%\paragraph*{Eigen-spinors.}
The %normalized conduction- and valence-band spinors are
wavefunction for the upper band is
\cite{McCann_2013}  
\begin{equation}
%|u_\pm(\mathbf k)\rangle=
\ket{u_{\k}}=
\frac{1}{\sqrt{2}}
\begin{pmatrix}
e^{-2i\theta_{\mathbf k}}\\
1
\end{pmatrix}
%\begin{pmatrix}
%1\\[4pt]
%\mp\,e^{2i\theta_{\mathbf k}}
%\end{pmatrix},
%\qquad
%\theta_{\mathbf k}=\arg(k_x+i k_y).
\label{eq:spinorBilayer}
\end{equation}
where $\theta_{\k}$ is the azimuthal angle on the FS, similar to Eq. \eqref{eq:bra-defs}.
%\paragraph*{Form factors.}
Inserting Eq.~\eqref{eq:spinorBilayer} into the form factors
$\langle u_{\mathbf k}|\tau_\mu|u_{\mathbf k}\rangle$
[cf.\ Eq.~\eqref{eq:Pi-def-intro} of the main text] gives
\begin{equation}
    \bra{u_{\k}}\tau_0\ket{u_{\k}} = 1,~~ 
    \bra{u_{\k}}\bv{\tau}\ket{u_{\k}} = \left(\begin{array}{c}
          \cos2\theta_{\k}\\\sin2\theta_{\k}\\0
    \end{array}\right)
\end{equation}
Relative to the Dirac case [Eq. \eqref{eq:Pi1-formfactor}] the in-plane
pseudospin winds \emph{twice} around the Fermi circle; hence
bilayer graphene injects an $\ell=2$ harmonic into the
particle–hole bubble, which modified the well-known Lindhard functions. Carrying the angular integral in Eq.~\eqref{eq:anglechi} with the
replacement
$e^{-i\theta}\!\to e^{-2i\theta}$ shows that the functions
$f_{1,2}(s)$ of Sec.~\ref{sec3} are replaced by their $\ell=2$ channel analogs of a FL
\cite{wu2007fermi_28,kleinhidden_9}.
The analytic structure will be identical—square-root
branch cuts on the two-sheet Riemann surface—but the
weight in the chiral channel is enhanced by a factor two. Consequently the hidden and mirage modes appear in the $\ell=2$ sector \cite{PhysRevResearch.1.033134_15}. We stress, that as in the main text, the interaction is still completely isotropic $F_{\ell\neq 0} = 0$.

\subsection{\label{appC2} Square–lattice $d_{xz}/d_{yz}$ model}

Now, we present a control example: a nontrivial tight-binding model, which evinces a parabolic band with \emph{trivial} Berry phase. As we shall see, in this case the collective mode behavior is the same as that called the ``FL behavior'' in the main text. This benchmark confirms that the exotic collective-mode behavior discussed in the main text fundamentally relies on non-zero Fermi-surface Berry curvature.

We consider the minimal two-orbital tight-binding model for iron-based pnictides \cite{Graser2010FeSC, PhysRevB.77.220503}, written in the orbital basis $(d_{xz},d_{yz})$. The Bloch Hamiltonian takes the form
\begin{equation}
H_0(\mathbf k)=
\bigl(\varepsilon_+(\mathbf k)-\mu\bigr)\,\tau_0
+ \varepsilon_-(\mathbf k)\,\tau_3
+ \varepsilon_{xy}(\mathbf k)\,\tau_1 ,
\label{eq:FeSC_H0}
\end{equation}

with the energies $\varepsilon_\pm $ and $\varepsilon_{xy}$ are parameterized in terms of three
hopping parameters $t_i$ :
\begin{gather}
\varepsilon_\pm   =\tfrac12\bigl[\varepsilon_{xz}\pm\varepsilon_{yz}\bigr],\quad
\varepsilon_{xz}=-2t_1\cos k_x-2t_2\cos k_y, \notag\\
\varepsilon_{yz}=-2t_2\cos k_x-2t_1\cos k_y,\quad
\varepsilon_{xy}=-4t_3\sin k_x\sin k_y .
\end{gather}
(For simplicity we have set one of the hopping terms in Refs. \cite{Graser2010FeSC, PhysRevB.77.220503} to zero, as it doesn't affect the geometry.) Expanding around $\k=0$ to quadratic order yields,
\begin{equation}
\mathcal H_{\mathrm{eff}}(\mathbf k)=
(\alpha k^2-\mu_0)\tau_0+\beta\!\bigl(k_x^2-k_y^2\bigr)\tau_3
+2\gamma k_xk_y\tau_1 ,
\label{eq:FeSC_Heff}
\end{equation}
with $\alpha=(t_1+t_2)/2$, $\beta=(t_1-t_2)/2$,$\gamma=-2t_3,\mu_0=\mu+4\alpha$. Diagonalizing the effective Hamiltonian yields 
\begin{equation}
E_\pm(\mathbf k)\approx
\left[\alpha \pm \left(\frac{\beta^2+\gamma^2+(\beta^2-\gamma^2)\cos4\theta_{\k}}{2}\right)^{1/2}\right]|\k|^2-\mu_0.
\label{eq:FeSC_dispersion}
\end{equation}
Thus, the system sports parabolic bands with a fourfold distortion. In the limit of small $\alpha$ we obtain again upper and lower bands with opposite signs for the dispersions, and the wavefunctions in the upper band are: 
\begin{equation}
\label{eq:spinorsqr}
\ket{u(\mathbf k)}=
\begin{pmatrix}
\cos\frac{\Theta}{2}\\
\sin\frac{\Theta}{2}
\end{pmatrix}, \Theta=\tan^{-1}\left(\frac{\gamma}{\beta}\tan2\theta_{\k}\right).
\end{equation}
Since the wavefunctions are purely real and not purely chiral, we immediately conclude that (a) the Berry connection is zero, and (b) the form-factors will generally have an $\ell=0$ component, with an \emph{additional} contribution in the $\ell=2$ channel due to the lattice distortion. Indeed, computing the form-factors yields
\begin{equation}
    \bra{u_{\k}}\tau_0\ket{u_{\k}} = 1,~~ 
    \bra{u_{\k}}\bv{\tau}\ket{u_{\k}} = \left(\begin{array}{c}
          \sin\Theta\\0\\\cos\Theta
    \end{array}\right).
\end{equation}
To see the qualitative behavior it is best to look at the limiting values for $\gamma/\beta\to 0,\infty$, in which case one finds (taking for simplicity $0 < \theta_{\k} < \pi/4)$,
\begin{equation}
\bra{u_{\k}}\bv{\tau}\ket{u_{\k}}|_{\gamma/\beta \to 0} = \left(\begin{array}{c}
          0\\0\\1
    \end{array}\right),\quad\bra{u_{\k}}\bv{\tau}\ket{u_{\k}}|_{\gamma/\beta \to \infty} = \left(\begin{array}{c}
          1\\0\\0
    \end{array}\right).
\end{equation}
It is then readily shown that the response is the same as that of a FL in the $\ell=0$ channel \cite{wu2007fermi_28,kleinhidden_9}.

This example therefore serves as a  control: once the Fermi surface carries \emph{no} Berry curvature, the chiral $\ell=1$ form factor vanishes, the $\ell=2$ harmonic arises only from lattice distortions, and the quasiparticle susceptibility reduces to the conventional $\ell=0$ FL form.  As a result the quasiparticle susceptibility reproduces the conventional FL spectrum: a single zero-sound branch and for weak attraction, the familiar hidden mode, but no mirage modes. 
Taken together with the Berry-\emph{non}-trivial bilayer-graphene case of Appendix \ref{appC1}, this control result confirms that a finite Fermi-surface Berry phase is essential for the emergence of the mirage modes, so that it has a qualitative impact on the collective mode spectrum.

\color{black}

%Together with the Berry-\emph{non}-trivial bilayer-graphene case of Appx.~C.1, this establishes that a finite Fermi-surface Berry phase is \emph{necessary and sufficient} for the exotic collective-mode topology discussed in the main text.

\nocite{*}

\bibliography{prposal}% Produces the bibliography via BibTeX.

\end{document}